\documentclass[11pt]{article}
\usepackage{pdfpages}

\usepackage{jmlr2e}
\usepackage{abstract}
\usepackage{array}
\usepackage{bbding}
\usepackage{multirow}
\usepackage{adjustbox}
\usepackage{bbm}
\usepackage{float}
\usepackage{amsfonts}
\usepackage{graphicx}
\graphicspath{{graphics/}}
\usepackage{subfigure}
\usepackage{float}
\usepackage{tabularx,booktabs}
\usepackage{multirow}
\newcolumntype{Y}{>{\centering\arraybackslash}X}
\usepackage[ruled,vlined,algo2e]{algorithm2e}
\usepackage{algorithm}
\usepackage{algorithmic}
\usepackage{natbib}

\usepackage{lscape}
\usepackage{comment}
\usepackage{bm}
\usepackage[title]{appendix}
\usepackage{longtable}
\usepackage{mathtools}
\usepackage{dsfont}
\usepackage{fancyhdr}

\usepackage{stackengine}
\usepackage{textcomp}
\usepackage{stfloats}
\usepackage{verbatim}
\usepackage{xcolor}
\definecolor{darkblue}{rgb}{0.0,0.5,0.5}
\definecolor{blue}{rgb}{0.0,0.0,1}

\usepackage[colorlinks]{hyperref}
\hypersetup{colorlinks,breaklinks,linkcolor=blue,urlcolor=blue,anchorcolor=blue,citecolor=blue}
\usepackage{cleveref}
\usepackage{booktabs,caption}
\usepackage{multirow}
\usepackage{lscape}
\usepackage{epstopdf}
\usepackage{lineno}
\usepackage{microtype}

\usepackage{graphicx}
\usepackage{subfigure}
\usepackage{subcaption}
\usepackage{caption}
\usepackage{booktabs} %
\usepackage{bbm}

\usepackage{amsfonts}
\usepackage{xurl}
\usepackage{stackengine}
\usepackage{tikz}
\usetikzlibrary{decorations.pathreplacing}
\usetikzlibrary{positioning,arrows.meta,quotes}
\usetikzlibrary{shapes,snakes}
\usetikzlibrary{bayesnet}
\tikzset{>=latex}
\tikzstyle{plate caption} = [caption, node distance=0, inner sep=0pt, below left=5pt and 0pt of #1.south]
\usepackage[normalem]{ulem}
\usepackage{multirow}
\usepackage{makecell}
\newcolumntype{Y}{>{\centering\arraybackslash}X}

\firstpageno{1}

\begin{document}

\title{Bayesian spatiotemporal modeling of passenger trip assignment in metro networks}

\author{\name Xiaoxu Chen \email xiaoxu.chen@mail.mcgill.ca\\
       \addr Department of Civil Engineering\\
       McGill University\\
       Montreal, QC  H3A 0C3, Canada
       \AND
       \name Alexandra M. Schmidt
       \email alexandra.schmidt@mcgill.ca \\
       \addr Department of Epidemiology, Biostatistics, and Occupational Health\\
       McGill University\\
       Montreal, QC H3A 1G1, Canada
       \AND
       \name Zhenliang Ma\email zhema@kth.se\\
       \addr Department of Civil and Architectural Engineering\\
       KTH Royal Institute of Technology\\
       Stockholm, 10044, Sweden
       \AND
       \name Lijun Sun\thanks{Corresponding author.} \email lijun.sun@mcgill.ca \\
       \addr Department of Civil Engineering\\
       McGill University\\
       Montreal, QC  H3A 0C3, Canada}

\editor{}

\maketitle

\begin{abstract}
Assigning passenger trips to specific network paths using automatic fare collection (AFC) data is a fundamental application in urban transit analysis. The task is a difficult inverse problem: the only available information consists of each passenger's total travel time and their origin and destination, while individual passenger path choices and dynamic network costs are unobservable, and behavior varies significantly across space and time. We propose a novel Bayesian hierarchical model to address this problem by jointly estimating dynamic network costs and passenger path choices while quantifying their uncertainty. Our model decomposes trip travel time into four components (access, in-vehicle, transfer, and egress), each modeled as a random walk over the day. To capture heterogeneous passenger behavior, we introduce a multinomial logit model with spatiotemporally varying coefficients. We manage the high dimensionality of these coefficients using kernelized tensor factorization with Gaussian process priors to effectively model complex spatiotemporal correlations. We develop a tailored and efficient Markov chain Monte Carlo (MCMC) algorithm for model inference. A simulation study demonstrates the method’s effectiveness in recovering the underlying model parameters. On a large-scale dataset from the Hong Kong Mass Transit Railway, our framework demonstrates better estimation accuracy over the benchmarks. The results reveal significant spatiotemporal variations in passenger preferences and provide robust uncertainty quantification, offering transit operators a powerful tool for enhancing service planning and operational management.
\end{abstract}


\section{Introduction}\label{sec:intro}

Metro systems are foundational to urban mobility, offering high-capacity transport that alleviates road congestion and promotes sustainable development. Central to their efficient operation is the \emph{passenger trip assignment} problem: determining the specific paths passengers take through the network. These assignments are critical inputs for optimizing nearly every aspect of transit service, from network design \citep{bagloee2011transit,bertsimas2021data} and train scheduling \citep{caprara2002modeling,jiang2022integrated} to fare policy \citep{bertsimas2020joint}.

Although modern Automatic Fare Collection (AFC) systems provide granular trip data, including origin-destination stations and tap-in/tap-out times, they do not record the actual path taken. This data gap makes trip assignment a challenging inverse problem. The difficulty is magnified by several factors: (1) \textbf{Unobservable path choices}: AFC data only yield total travel time, not the sequence of links and transfers that constitute a path. (2) \textbf{Latent network costs}: Key attributes influencing path choice, such as in-vehicle and transfer times, are themselves unobserved and dynamic. (3) \textbf{Behavioral heterogeneity}: Passenger preferences vary significantly across different locations and times of day. (4) \textbf{Pervasive uncertainty}: All estimated quantities, from path choices to model parameters, are subject to uncertainty, necessitating a probabilistic approach. (5) \textbf{High dimensionality}: The combination of dynamic costs and spatiotemporal behavioral parameters creates a high-dimensional estimation problem that is computationally intensive for large metro networks.

Many studies have sought to solve the trip assignment problem, but significant challenges remain unsolved. Early studies, extensively reviewed by \citet{liu2010transit}, relied on strong simplifying assumptions because of limited data. Later studies improved model calibration by utilizing preference survey data \citep{lam2002transit,nazem2011demographic}, though they suffered from small sample sizes and biases. The emergence of large-scale AFC data marked a significant advancement, enabling data-driven approaches that infer passenger paths. Many studies inferred paths by matching AFC data to train operations data \citep{kusakabe2010estimation,zhao2016estimation,zhu2017probabilistic,zhu2021passenger}. Such ``matching'' methods often rely on deterministic assumptions about operational schedules or simplified passenger behavior, overlooking crucial aspects like behavioral heterogeneity and operational uncertainty. More recent work has focused on the joint estimation of both path choices and network cost attributes as latent variables. The integrated Bayesian framework proposed by \cite{sun2015integrated} was a landmark in this area, demonstrating that both could be estimated simultaneously from AFC data alone. This work inspired subsequent studies that incorporated additional real-world factors, such as crowdedness \citep{xu2018learning}, backward-travel behaviors \citep{yu2020data}, and optimization-based methods \citep{mo2023ex}. These methods have \textit{partially} addressed the challenges posed by unobservable passenger paths (Challenge 1) and latent network costs (Challenge 2) by jointly estimating these quantities from AFC data. Some approaches have also provided a probabilistic quantification of associated uncertainties (Challenge 4). However, current frameworks remain fundamentally \textit{static}, overlooking crucial temporal variations in network costs and the spatiotemporal heterogeneity in passenger path choice behavior (Challenges 2 and 3). Modeling spatiotemporal variations can achieve more accurate path choice estimation and robust uncertainty quantification (Challenge 4). Accounting for these spatiotemporal variations will inherently lead to a high-dimensional estimation problem (Challenge 5), posing computational complexities that previous methods have not adequately addressed.

To overcome these challenges, we propose a novel Bayesian framework that \textit{jointly} estimates dynamic network costs and spatiotemporal passenger path choices. Our approach models the observed trip travel time as a noisy linear combination of its constituent costs, i.e., access, in-vehicle, transfer, and egress, which are themselves modeled as dynamic random walks. To capture the evolution of passenger behavior, we introduce a multinomial logit (MNL) model with spatiotemporally varying coefficients. We manage the high dimensionality of these coefficients by employing kernelized tensor factorization, which parsimoniously captures complex spatiotemporal patterns via Gaussian process (GP) priors. Inference is performed using a tailored and efficient Markov chain Monte Carlo (MCMC) algorithm.

From a statistical point of view, the choice model is a spatiotemporally varying coefficient model in the spirit of \citet{gelfand2003spatial}. Two features distinguish it from that literature: the coefficients enter a discrete-choice likelihood whose covariates, the path attributes, are themselves latent states; and the $n\times T$ coefficient fields are regularized through the low-rank kernelized tensor representation of \citet{JMLR:v18:16-362} and \citet{lei2024scalable} rather than through a full Gaussian process. The attributes follow a conditionally linear Gaussian state-space model, and the two components are linked by a modularized Gibbs sampler.

We demonstrate the effectiveness and robustness of our framework through an extensive simulation study and an application to a large-scale dataset from the Hong Kong Mass Transit Railway (MTR). Our model yields more accurate estimates than the benchmarks, underscoring the critical importance of incorporating spatiotemporal dynamics. The resulting parameter analyses and visualizations of passenger flows offer powerful, interpretable tools for transit management and operations.

The main contributions of this paper are:
\begin{enumerate}
    \item A Bayesian framework that provides a dynamic and accurate representation of passenger behavior by jointly estimating time-varying network costs and path choices.
    \item An MNL choice model with spatiotemporally varying coefficients that captures the evolving nature of passenger preferences for in-vehicle and transfer costs.
    \item The use of kernelized tensor decomposition to efficiently manage the high-dimensional parameter space of the choice model, enabling the capture of complex spatiotemporal variations.
    \item An efficient MCMC-based inference algorithm, using forward filtering backward sampling (FFBS) and elliptical slice sampling, that makes large-scale estimation computationally feasible.
    \item A comprehensive model validation on simulated data and real-world AFC records, demonstrating the model's ability to accurately assign passenger flows and provide interpretable insights with robust uncertainty quantification.
\end{enumerate}

The remainder of this paper is organized as follows. Section~\ref{sec:problem} formally defines the trip assignment problem. Section~\ref{sec:model} describes the proposed Bayesian framework. Section~\ref{sec:inference} develops the inference procedure. Section~\ref{sec:exp} presents the simulation study and the application to the Hong Kong MTR data. Finally, Section~\ref{sec:con} concludes and discusses future research directions.

\section{Overview of 
 the Problem}\label{sec:problem}

We begin by formally defining the passenger trip assignment problem. Our goal is to infer the specific paths passengers take through a metro network, given only their entry and exit times and locations.

\subsection{Notation and a Model for Trip Travel Time}
Consider a metro network with $n$ stations, $l$ directed in-vehicle links, and $s$ transfer links. For any origin-destination (O-D) pair $(o,d)$, there exists a set of feasible paths, $R_{od} = \{r_{od}^k\}_{k=1}^{|R_{od}|}$. When multiple paths exist ($|R_{od}| > 1$), passengers choose a path based on various factors. Our observed data, derived from the AFC system, consist of a set of trip travel times. Each observation $y_{odt}^m$ is the travel time for trip $m$, taken by a passenger traveling from origin station $o$ to destination station $d$, starting within time interval $t\in\{1,\ldots,T\}$. Our primary objective is to infer the latent path choice, $z_{odt}^m \in \{1, \dots, |R_{od}|\}$, for each trip observation.

\begin{figure*}[!t]
\centering
\includegraphics[width=0.98\textwidth]{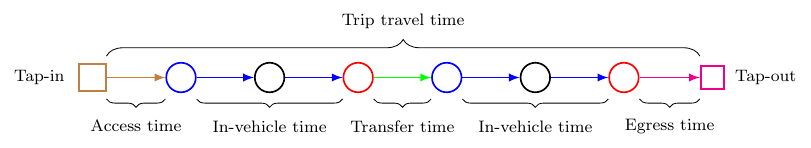}
\caption{The components of trip travel time. The brown square marks the tap-in gate where the trip begins, the blue circle indicates the point of boarding the train, the black circle represents the state of passengers being onboard the train, the red circle signifies alighting from the train, and the pink square marks the tap-out gate upon exit.}
\label{fig:components}
\end{figure*}

To model the observed travel time $y_{odt}^m$, we first decompose it into four constituent components, as illustrated in Fig.~\ref{fig:components}: (1)~\textbf{access time}, from entry tap-in to boarding; (2)~\textbf{in-vehicle time}, spent on the train; (3)~\textbf{transfer time}, the duration between alighting and boarding connecting trains; and (4)~\textbf{egress time}, from alighting to exit tap-out. Because network conditions change over the day, we index these costs by time interval and collect them in a single network cost vector $\boldsymbol{x}_t\in\mathbb{R}^c$, $c=2l+n+s$, which stacks $l$ access costs, $l$ in-vehicle costs, $s$ transfer costs, and $n$ egress costs. Access is attached to the first in-vehicle link of a path and is therefore direction-specific (from origin station $o$ and destination station $d$), which is why it has the same dimension as the in-vehicle links.

 For a given path choice $z_{odt}^m = k$, the observed travel time $y_{odt}^m$ can be expressed as a linear function of the network costs at time $t$:
\begin{equation}
y_{odt}^m = \boldsymbol{A}_{od}^k \boldsymbol{x}_t + \varepsilon_{odt}^m,
\label{eq:linear}
\end{equation}
where $\boldsymbol{A}_{od}^k$ is the binary routing row vector (of dimension $1\times c$) that selects the access, in-vehicle, transfer and egress components of path $r_{od}^k$ and $\varepsilon_{odt}^m$ is a zero-mean error term.

\subsection{Core Challenges}
The formulation in Eq.~\eqref{eq:linear} reveals the two fundamental challenges of the passenger trip assignment problem. First, this is a severely ill-posed inverse problem. In practice, only the total travel time $y_{odt}^m$ is observed. The two key explanatory variables, the latent path choice $z_{odt}^m$ and the dynamic network costs $\boldsymbol{x}_t$, are unobserved and must be inferred simultaneously from the data. Second, a credible assignment requires an accurate model of passenger path choice. We employ an MNL model \citep{mcfadden1974conditional} to capture the trade-offs passengers make between factors like in-vehicle and transfer time. Critically, passenger preferences are not static; they exhibit significant spatiotemporal heterogeneity. For example, passengers starting from central stations during peak hours may weigh transfer times differently than those traveling from peripheral stations during off-peak periods. Modeling these evolving preferences is essential for realism, but estimating the resulting high-dimensional, spatiotemporally varying parameters is a formidable challenge, especially in O-D pairs or time intervals with sparse data.

\section{A Spatiotemporal Bayesian Framework}\label{sec:model}

To address the challenges outlined in Section~\ref{sec:problem}, we develop a Bayesian hierarchical framework that is conditionally linear Gaussian: given the passengers' path choices, the observation and state-transition equations form a linear Gaussian state-space model. This framework allows us to jointly infer the latent network states (costs) and passenger path choices. The framework consists of three main components: an observation model linking latent network costs and path choices to observed trip travel times, a state-transition model describing the evolution of network costs, and a path-choice model characterizing spatiotemporal variation in passenger preferences.

\subsection{Observation Model: A Heteroscedastic Model for Trip Travel Times}
The observation model connects the latent network costs $\boldsymbol{x}_t$ to the observed trip data $\boldsymbol{Y}$. As formulated in Eq.~\eqref{eq:linear}, the travel time $y_{odt}^m$ for a given path $k$ is a linear function of the costs, $y_{odt}^m = \boldsymbol{A}_{od}^k\boldsymbol{x}_{t} + \varepsilon_{odt}^m$. A simple homoscedastic assumption for the error term ($\varepsilon_{odt}^m \sim \mathcal{N}(0, \sigma^2)$) is unrealistic, as longer and more complex paths are inherently more variable.

To capture this heterogeneity in a parsimonious way, we model the variance of the error term as the sum of component-specific variances along the path. A natural starting point is to assign separate error variances to the four travel-time components: access, in-vehicle, transfer, and egress. However, each observed trip always contains exactly one access component and one egress component. As a result, under this four-parameter formulation, the likelihood depends on the access and egress variances only through their sum, $\sigma_a^2 + \sigma_e^2$, so these two variances are not separately identifiable from AFC trip travel times alone. We therefore reparameterize the model using a combined {access-egress} error {variance} $\sigma_{ae}^2 = \sigma_a^2 + \sigma_e^2$, while retaining separate {variance} parameters for in-vehicle and transfer variability.

For a feasible path $r_{od}^k$, let $n_{h,od}^k$ and $n_{u,od}^k$ denote the numbers of in-vehicle links and transfer links on that path, respectively, and define the variance-design row vector $\boldsymbol{c}_{od}^k = \left(1, n_{h,od}^k, n_{u,od}^k\right)$. We then define the error model as:
\begin{equation}
    \varepsilon_{odt}^m \mid \boldsymbol{x}_t, \boldsymbol{\sigma}^2, z_{odt}^m=k \sim \mathcal{N}\left(0, \boldsymbol{c}_{od}^k \boldsymbol{\sigma}^2\right),
    \label{eq:error}
\end{equation}
where $\boldsymbol{\sigma}^2 =
\left(\sigma_{ae}^2, \sigma_h^2, \sigma_u^2\right)^\top
\in \mathbb{R}_+^3$ is the vector of variance parameters.
Specifically, $\sigma_{ae}^2$ represents
the combined access-egress error variance, while
$\sigma_h^2$ and $\sigma_u^2$ represent the error
variances per in-vehicle link and per transfer link,
respectively. The first element of $\boldsymbol{c}_{od}^k$ corresponds to the combined access-egress variability, while the second and third elements count the in-vehicle and transfer components along the path. Combining these elements gives the full observation equation:
\begin{equation}
    y_{odt}^m \mid \boldsymbol{x}_t, \boldsymbol{\sigma}^2, z_{odt}^m=k \sim \mathcal{N}\left(\boldsymbol{A}_{od}^{k}\boldsymbol{x}_{t}, \boldsymbol{c}_{od}^k\boldsymbol{\sigma}^2\right).
    \label{eq:linear_Gaussian}
\end{equation}

The parsimonious variance formulation offers two key advantages: (1) it avoids the need to estimate path-specific variances, which would be infeasible in large-scale networks; and (2) it allows variance to naturally scale with path length and composition, capturing the greater uncertainty associated with longer or more complex trips. In particular, the combined access-egress term captures the unavoidable entry/exit variability of a trip, while the in-vehicle and transfer terms increase with the number of corresponding path components. This three-parameter specification resolves the non-identifiability of separate access and egress variances while still capturing the main sources of residual uncertainty in passenger travel.  

\subsection{State Transition Model: A Random Walk for Dynamic Network Costs}
To capture the dynamic nature of the metro system, we model the evolution of the network cost vector $\boldsymbol{x}_t$ using a first-order random walk. This represents a simple yet effective assumption that costs in a given time interval are a perturbation of the costs from the previous interval. The state transition is defined as:
\begin{equation}\label{eq:random_walk}
\begin{aligned}
    \boldsymbol{x}_t \mid \boldsymbol{x}_{t-1} &\sim \mathcal{N}\left(\boldsymbol{x}_{t-1}, \text{diag}(\boldsymbol{\tau}^2)\right), \quad t=2,\ldots,T,  \\ 
    \boldsymbol{x}_1 &\sim \mathcal{N}(\boldsymbol{m}_0, \boldsymbol{P}_0),
\end{aligned}
\end{equation}
where the process noise $\boldsymbol{\tau}^2 \in \mathbb{R}^c_+$ governs the magnitude of change between intervals, and $(\boldsymbol{m}_0, \boldsymbol{P}_0)$ are the prior mean and covariance of the initial state.

\subsection{Path Choice Model: MNL with Spatiotemporally Varying Coefficients}
The MNL model is widely adopted for modeling metro passengers' path choices. Under the MNL model \citep{mcfadden1974conditional}, each feasible path $r_{od}^k$ between an O-D pair $(o,d)$ is chosen with probability proportional to $\exp(V_{od}^k)$, where the deterministic utility $V_{od}^k$ typically takes a linear form with respect to path attributes, such as in-vehicle travel time and transfer time; the coefficients represent passengers' sensitivities to these attributes \citep{sun2015integrated}.

Conventional MNL implementations for metro route choices assume constant utility coefficients across all passenger trips, implicitly treating passenger behavior as homogeneous. However, passenger path choices in reality are heterogeneous, influenced by factors such as trip purpose, socio-demographic characteristics, and contextual conditions, most of which are unavailable in AFC data. Ignoring this heterogeneity and applying universal parameters can lead to biased estimates, misleading inference, and limited predictive capability. Moreover, the standard MNL imposes the independence of irrelevant alternatives (IIA) property, which can be restrictive in metro networks where feasible paths often share common links. Although allowing utility coefficients to vary does not eliminate the IIA property, it can capture some of the systematic heterogeneity in path preferences across space and time that would otherwise be overlooked by a fixed-coefficient model.

To better capture the underlying behavioral variability, we propose a spatiotemporal MNL model by allowing the utility coefficients associated with in-vehicle time and transfer time to vary smoothly across both the origin station and the time of day. Specifically, we define the deterministic utility $V_{odt}^k$ for a path $r_{od}^k$ at origin station $o$ during time interval $t$ as follows:
\begin{equation}
V_{odt}^{k} = \theta_{ot}\sum_{a\in \mathcal{I}_{od}^k}{x}_{t}^a + \phi_{ot}\sum_{a\in \mathcal{T}_{od}^k}{x}_{t}^a,
\label{eq:utility}
\end{equation}
where $\theta_{ot}$ and $\phi_{ot}$ represent the spatially and temporally varying sensitivities to in-vehicle time and transfer time, respectively. Here, $\mathcal{I}_{od}^k$ and $\mathcal{T}_{od}^k$ denote the sets of in-vehicle links and transfer links on the path, respectively, and ${x}_{t}^a$ represents the entry of $\boldsymbol{x}_t$ associated with link $a$. 

Under this specification, the probability that path $k$ is chosen at time interval $t$ is expressed by the standard logit formula:
\begin{equation}
p(z_{odt}^m = k \mid \boldsymbol{x}_t, \theta_{ot}, \phi_{ot}) = \frac{\exp(V_{odt}^k)}{\sum_{k'=1}^{|R_{od}|}\exp(V_{odt}^{k'})},
\label{eq:choice}
\end{equation}
where $z_{odt}^m$ denotes the latent path choice for a passenger traveling from $o$ to $d$ during time interval $t$, and $|R_{od}|$ is the total number of feasible paths for the O-D pair $(o,d)$. This approach provides a flexible, behaviorally realistic framework that accounts explicitly for spatial and temporal heterogeneity in passenger preferences.

A natural method to introduce spatiotemporal variations is to use GPs to model the varying parameters $\boldsymbol{\Theta} = \{\theta_{ot}\} \in \mathbb{R}^{n \times T}$ and $\boldsymbol{\Phi} = \{\phi_{ot}\} \in \mathbb{R}^{n \times T}$, leveraging relationships between nearby stations and adjacent periods to improve the accuracy of the estimates. For example, one can assume:
$\text{vec}\left(\boldsymbol{\Theta}\right) \sim\mathcal{N}\left(\textbf{0}_{nT},\boldsymbol{K}_T\otimes\boldsymbol{K}_S\right)$ and $\text{vec}\left(\boldsymbol{\Phi}\right) \sim\mathcal{N}\left(\textbf{0}_{nT},\boldsymbol{K}_T\otimes\boldsymbol{K}_S\right)$, where $\boldsymbol{K}_S\in
\mathbb{R}^{n\times n}$ and $\boldsymbol{K}_T\in
\mathbb{R}^{T\times T}$ are the spatial and temporal covariance matrices, respectively. However, the dimension of $\text{vec}\left(\boldsymbol{\Theta}\right)$ and $\text{vec}\left(\boldsymbol{\Phi}\right)$ is $nT$, which presents a significant computational challenge for parameter estimation. This high dimensionality makes even advanced methods like elliptical slice sampling struggle to achieve convergence due to the large parameter space. To address this, we propose to model $\boldsymbol{\Theta}$ and $\boldsymbol{\Phi}$ as a single third-order parameter tensor $\boldsymbol{\mathcal{F}} \in \mathbb{R}^{2 \times n \times T}$, where $\boldsymbol{\mathcal{F}}\left(1,:,:\right)=\boldsymbol{\Theta}$ and $\boldsymbol{\mathcal{F}}\left(2,:,:\right)=\boldsymbol{\Phi}$, and apply a kernelized CANDECOMP/PARAFAC (CP) decomposition \citep{kolda2009tensor,lei2024scalable}:
\begin{equation}
    \boldsymbol{\mathcal{F}} = \boldsymbol{\mathcal{Q}} + \sum_{r=1}^R \boldsymbol{u}_r \circ \boldsymbol{v}_r \circ \boldsymbol{w}_r,
    \label{eq:tensor-1}
\end{equation}
where $R$ is the CP rank, $\circ$ denotes the outer product, ${\boldsymbol{u}}_r\in\mathbb{R}^2$, ${\boldsymbol{v}}_r\in \mathbb{R}^n$, and ${\boldsymbol{w}}_r\in\mathbb{R}^T$ are the latent factor vectors corresponding to choice (parameter) type, station, and time, respectively. The tensor $\boldsymbol{\mathcal{Q}} \in \mathbb{R}^{2 \times n \times T}$ represents the baseline effect, defined such that $\boldsymbol{\mathcal{Q}}(1,:,:)=q_1\cdot\boldsymbol{1}_{n\times T}$ and $\boldsymbol{\mathcal{Q}}(2,:,:)=q_2\cdot\boldsymbol{1}_{n\times T}$. Let ${\boldsymbol{U}}=\left({\boldsymbol{u}}_1,\ldots,{\boldsymbol{u}}_R\right)\in \mathbb{R}^{2\times R}$, ${\boldsymbol{V}}=\left({\boldsymbol{v}}_1,\ldots,{\boldsymbol{v}}_R\right)\in \mathbb{R}^{n\times R}$, and ${\boldsymbol{W}}=\left({\boldsymbol{w}}_1,\ldots,{\boldsymbol{w}}_R\right)\in \mathbb{R}^{T\times R}$ be the factor matrices. This decomposition significantly reduces the number of parameters from $2nT$ to $(2+n+T)R+2$, where $R$ is typically much smaller than $n$ and $T$, enabling efficient estimation and scalable inference while preserving the ability to capture complex spatiotemporal passenger choice behavior.

To ensure the estimated coefficients are smooth and correlated across space and time, we place GP priors on the spatial and temporal factors following  \cite{lei2024scalable}:
\begin{equation}
\begin{aligned}
    \boldsymbol{v}_r &\sim \mathcal{N}(\boldsymbol{0}_{n}, \boldsymbol{K}_S), \quad r=1,\ldots,R, \\
    \boldsymbol{w}_r &\sim \mathcal{N}(\boldsymbol{0}_{T}, \boldsymbol{K}_T), \quad r=1,\ldots,R.
\end{aligned} \label{eq:GP_vr}
\end{equation}

The spatial covariance matrix $\boldsymbol{K}_S$ is constructed from a diffusion kernel on the station adjacency graph \citep{smola2003kernels}, $\boldsymbol{K}_S=\text{expm}\left(-\alpha\mathcal{L}_{\text{norm}}\right)$, where $\text{expm}\left(\cdot\right)$ is the matrix exponential, $\alpha$ is a scaling parameter, and $\mathcal{L}_{\text{norm}}=\boldsymbol{D}^{-\frac{1}{2}}\left(\boldsymbol{D}-\boldsymbol{J}\right)\boldsymbol{D}^{-\frac{1}{2}}$ is the normalized graph Laplacian based on the network's adjacency ($\boldsymbol{J}$) and degree ($\boldsymbol{D}$) matrices. For temporal covariance $\boldsymbol{K}_T$, we use the squared-exponential kernel \citep{rasmussen2006gaussian}, $[\boldsymbol{K}_T]_{tt'} = \sigma_T^2\exp\left(-\frac{\left(t-t'\right)^2}{2\ell_T^2}\right)$, with length-scale $\ell_T$ and variance $\sigma_T^2$. These priors allow the model to borrow strength across nearby stations and adjacent time intervals, leading to robust estimation. The squared-exponential kernel encodes the assumption that sensitivities change smoothly over the day; sharper short-term fluctuations in network conditions are absorbed by the latent costs $\{\boldsymbol{x}_t\}$ rather than by the temporal factors. 

For factor matrix ${\boldsymbol{U}}$, we assume the prior ${\boldsymbol{u}}_r \sim \mathcal{N}\left(\textbf{0}_{2},\boldsymbol{K}_U\right)$. Further, we impose an inverse Wishart prior on $\boldsymbol{K}_U$:
\begin{equation}
    \boldsymbol{K}_U \sim \mathcal{W}^{-1}\left(\boldsymbol{\Omega}_0,\nu_0\right), \label{eq:cov prior} \\
\end{equation}
where $\boldsymbol{\Omega}_0$ is the scale matrix and $\nu_0$ represents degrees of freedom. 

Once the factor matrices ${\boldsymbol{U}}$, ${\boldsymbol{V}}$, and ${\boldsymbol{W}}$ are estimated, the parameter tensor $\boldsymbol{\mathcal{F}}$ can be reconstructed using the CP decomposition. Subsequently, the spatiotemporal path choice parameters $\boldsymbol{\Theta}$ and $\boldsymbol{\Phi}$ are obtained as slices of the reconstructed tensor. Although the CP decomposition factor matrices $\boldsymbol{U}$, $\boldsymbol{V}$, and $\boldsymbol{W}$ are not individually identifiable due to permutation and scaling indeterminacies inherent in tensor factorization, the parameter tensor $\boldsymbol{\mathcal{F}}$, which is the only quantity entering the likelihood, is invariant to them. As shown in the tensor regression framework proposed by \cite{JMLR:v18:16-362}, the tensor $\boldsymbol{\mathcal{F}}$ remains uniquely determined even without imposing explicit identifiability restrictions on the individual factor matrices. Therefore, we avoid enforcing unnecessary identifiability constraints on $\boldsymbol{U}$, $\boldsymbol{V}$, and $\boldsymbol{W}$, as our inference is solely based on the  tensor parameter $\boldsymbol{\mathcal{F}}$.

\subsection{Prior Specifications}\label{sec:prior}
We complete the model by specifying priors for the remaining parameters. For the error variances, we use Gaussian priors on the log-transformed parameters: $\log{\left(\sigma_{ae}^2\right)}\sim \mathcal{N}\left(\mu_{\sigma_{ae}^2},\sigma^2_{aep}\right)$, $\log{\left(\sigma_h^2\right)}\sim\mathcal{N}\left(\mu_{\sigma_h^2},\sigma^2_{h p}\right)$, and $\log{\left(\sigma_u^2\right)}\sim \mathcal{N}\left(\mu_{\sigma_u^2},\sigma^2_{up}\right)$, where we set hyperparameters $\mu_{\sigma_{ae}^2}=\mu_{\sigma_h^2}=\mu_{\sigma_u^2}=0.5$ and $\sigma_{aep}^2=\sigma_{hp}^2=\sigma_{up}^2=6.25$. These choices reflect weakly informative priors, expressing a mild belief that variation in travel-time components is limited but non-negligible. The prior variances allow moderate flexibility without dominating the inference. For the baseline effects $q_1$ and $q_2$, we use Gaussian priors as: $q_1\sim \mathcal{N}\left(\mu_{q_1},\sigma^2_{q_1}\right)$ and $q_2\sim \mathcal{N}\left(\mu_{q_2},\sigma^2_{q_2}\right)$, where we set $\mu_{q_1}=\mu_{q_2}=0$ and $\sigma^2_{q_1}=\sigma^2_{q_2}=0.1$. These priors reflect weak prior knowledge about the average scale of the latent utility parameters and allow deviations to be primarily learned from the data. We set $\ell_T=3$ for the length-scale of the temporal kernel function to allow for moderate temporal smoothing across a few hours, and the variance to $\sigma_T^2=1$; for the diffusion kernel, we set $\alpha=0.2$ to allow for spatial correlation without excessive shrinkage toward homogeneity. For the hyperparameters in the inverse Wishart prior, we set $\boldsymbol{\Omega}_0=\boldsymbol{I}_{2}$ and $\nu_0=5$. This defines a weakly informative prior with a small number of degrees of freedom, offering mild regularization while retaining flexibility in capturing latent structure in the decomposition. {The hyperparameters in these priors are chosen to be weakly informative and to impose only mild smoothing or regularization. In our experiments, posterior inference was qualitatively stable under reasonable alternative choices of these hyperparameters, suggesting that the results are primarily driven by the data rather than by the specific prior settings.}

\section{Inference Procedure}\label{sec:inference}
We now describe the procedure for estimating the parameters of our model given the observed data. We begin by specifying the likelihood for a single observed trip. For an individual observation $y_{odt}^m$, the likelihood is $p\left(y_{odt}^m\mid \boldsymbol{x}_t,z_{odt}^m,\boldsymbol{\sigma}^2\right)$. Let $\boldsymbol{Y} = \{y_{odt}^m\}$  denote the full set of observed trip travel times, for $m=1,\cdots, M_{odt}$, $t=1,\cdots, T$, and $(o,d)$ that are within $B$, the set of all possible O-D pairs; and $\boldsymbol{Z}=\{z_{odt}^m\}$ be the set of all latent path choice variables. We define the collection of model parameters as 
$\boldsymbol{\Gamma} = \left\{\{\boldsymbol{x}_t\}_{t=1}^T, \boldsymbol{\sigma}^2, \boldsymbol{U}, \boldsymbol{V}, \boldsymbol{W}, \boldsymbol{K}_U, q_1, q_2\right\}$. The process-noise variances $\boldsymbol{\tau}^2$ and the initial-state moments $(\boldsymbol{m}_0,\boldsymbol{P}_0)$ are treated as fixed; the values used are given in Section~\ref{sec:exp}.

Following Bayes' theorem, the joint posterior distribution of the parameters and latent variables, $p(\boldsymbol{\Gamma}, \boldsymbol{Z} \mid \boldsymbol{Y})$, is proportional to the product of the likelihood and the prior distribution, that is,
\begin{equation}\label{eq:posterior}
\begin{aligned}
    p\left(\boldsymbol{\Gamma}, \boldsymbol{Z} \mid \boldsymbol{Y}\right)  \propto{} &  \prod_{\left(o,d\right)\in B}\prod_{t=1}^T\prod_{m=1}^{M_{odt}} p\left(y_{odt}^m\mid \boldsymbol{x}_t, z_{odt}^m, \boldsymbol{\sigma}^2\right)\, p\left(z_{odt}^m \mid\boldsymbol{x}_t, \boldsymbol{\Theta}, \boldsymbol{\Phi}\right)   \\ 
    & \times p\left(\{\boldsymbol{x}_t\} \mid \boldsymbol{\tau}^2\right)\, p\left(\boldsymbol{U}\mid\boldsymbol{K}_U\right) p\left(\boldsymbol{K}_U\right) p\left(\boldsymbol{V}\right) p\left(\boldsymbol{W}\right) p\left(q_1\right) p\left(q_2\right) p\left(\boldsymbol{\sigma}^2\right),
\end{aligned}
\end{equation}%
where $\boldsymbol{\Theta}$ and $\boldsymbol{\Phi}$ are the deterministic functions of $(\boldsymbol{U},\boldsymbol{V},\boldsymbol{W},q_1,q_2)$ given by Eq.~\eqref{eq:tensor-1}, and the priors are as specified in Section~\ref{sec:model}. We assume prior independence among the components of $\boldsymbol{\Gamma}$.

The posterior is high-dimensional and analytically intractable due to the hierarchical structure, non-conjugacy, and latent discrete variables. Consequently, we develop an efficient MCMC algorithm based on Gibbs sampling. We divide the parameter space into four blocks, which are updated in turn:
\begin{enumerate}
    \item The time-varying network costs, $\{\boldsymbol{x}_t\}$.
    \item The error variances for the travel time error model, $\{\sigma_{ae}^2,\sigma_h^2,\sigma_u^2\}$.
    \item The parameters of the path choice model, which are governed by the tensor factors ($\boldsymbol{U}, \boldsymbol{V}, \boldsymbol{W}$), their covariance $\boldsymbol{K}_U$, and baseline effects ($q_1, q_2$).
    \item The latent path choice outcomes, $\boldsymbol{Z}$.
\end{enumerate}
Blocks 2-4 are updated by kernels that leave the joint posterior in Eq.~\eqref{eq:posterior} invariant: exact draws, slice sampling, or elliptical slice sampling with $\boldsymbol{Z}$ integrated out. Block 1 is updated from a modularized conditional that omits the choice-model term  (Section~\ref{sec:sample_x}), so the overall sampler is an approximation to Eq.~\eqref{eq:posterior}. In the following subsections, we describe the update for each block.

\subsection{Sampling the Time-Varying Network Costs $\{\boldsymbol{x}_t\}$}
\label{sec:sample_x}
Under the joint posterior in Eq.~\eqref{eq:posterior}, the full conditional distribution of the network costs is
\begin{equation}
    p\left(\{\boldsymbol{x}_t\} \mid \cdot\right) \propto p\left(\{\boldsymbol{x}_t\} \mid \boldsymbol{\tau}^2\right) \prod_{\left(o,d\right)\in B}\prod_{t=1}^T\prod_{m=1}^{M_{odt}} p\left(y_{odt}^m\mid \boldsymbol{x}_t, z_{odt}^m, \boldsymbol{\sigma}^2\right) p\left(z_{odt}^m \mid\boldsymbol{x}_t, \boldsymbol{\Theta}, \boldsymbol{\Phi}\right).
    \label{eq:x_full}
\end{equation}
The last factor depends on $\boldsymbol{x}_t$ through the utilities $V_{odt}^k$ in Eqs.~\eqref{eq:utility}-\eqref{eq:choice} and is not Gaussian in $\boldsymbol{x}_t$. We therefore use a modularized (``cut'') update \citep{bayarri2009modularization,plummer2015cuts}: when updating $\{\boldsymbol{x}_t\}$, we omit the MNL factor and draw from
\begin{equation}
    \tilde{p}\left(\{\boldsymbol{x}_t\} \mid \cdot\right) \propto p\left(\{\boldsymbol{x}_t\} \mid \boldsymbol{\tau}^2\right) \prod_{\left(o,d\right)\in B}\prod_{t=1}^T\prod_{m=1}^{M_{odt}} p\left(y_{odt}^m\mid \boldsymbol{x}_t, z_{odt}^m, \boldsymbol{\sigma}^2\right).
    \label{eq:x_cut}
\end{equation}
Equivalently, the path attributes entering the choice model, $\sum_{a\in \mathcal{I}_{od}^k}x_t^a$ and $\sum_{a\in \mathcal{T}_{od}^k}x_t^a$, are held fixed at their current values within this step, so the MNL factor is constant in the $\boldsymbol{x}_t$ being drawn. The network costs are thus informed only by the travel time likelihood and the random-walk prior. The dependence of the choice probabilities on $\boldsymbol{x}_t$ is still used when updating the choice parameters and the path choices, both of which condition on the current draw of $\{\boldsymbol{x}_t\}$; only the feedback from the choice model to the network costs is removed. In the language of modular Bayesian inference \citep{lunn2009combining,bayarri2009modularization,jacob2017better}, the travel-time model of Eqs.~\eqref{eq:linear_Gaussian}-\eqref{eq:random_walk} and the choice model of Eqs.~\eqref{eq:utility}-\eqref{eq:choice} are two modules that share the network costs. The choice model still informs the costs indirectly, through the path assignments $\boldsymbol{Z}$ that determine the routing matrices, but its likelihood is not allowed to pull the costs toward values that would rationalize the assigned path shares. Because the kernel for Block 1 targets Eq.~\eqref{eq:x_cut} whereas the kernels for Blocks 2-4 target Eq.~\eqref{eq:posterior}, the conditional distributions used by Algorithm~\ref{al:overall} are not mutually compatible. Algorithm~\ref{al:overall} is therefore a pseudo-Gibbs sampler: its stationary distribution approximates Eq.~\eqref{eq:posterior} but has no closed form \citep[see][for the analogous property of cut models]{plummer2015cuts}.

We adopt this approximation for two reasons. First, Eq.~\eqref{eq:x_cut} allows the whole trajectory $\{\boldsymbol{x}_t\}$ to be drawn jointly and exactly by Forward Filtering Backward Sampling (FFBS). An exact update of Eq.~\eqref{eq:x_full} is possible in principle, for example by using the FFBS draw as a Metropolis-Hastings proposal with acceptance ratio $\prod_{(o,d),t,m} p(z_{odt}^m \mid \boldsymbol{x}_t^{\text{new}}, \cdot)/p(z_{odt}^m \mid \boldsymbol{x}_t^{\text{old}}, \cdot)$, in which the Gaussian factors cancel exactly. However, this ratio is a product over millions of trips and the entire trajectory is proposed at once, so its acceptance rate would typically be very low; proposing one interval at a time would raise the acceptance rate, but at the cost of the joint draw of the trajectory. Second, cutting the feedback prevents the behavioral choice model, whose MNL form (including the IIA property) is the component most likely to be misspecified, from distorting the network costs, which have a direct interpretation as travel times; confining the consequences of a misspecified module to that module is the main argument for cutting feedback in \citet{jacob2017better}. Empirically, in the simulation study of Section~\ref{sec:simulation}, where the choice model is correctly specified and the cut can only discard information, the modularized sampler still recovers $\{\boldsymbol{x}_t\}$ accurately, with credible intervals that cover the true trajectories (Fig.~\ref{fig:sim_parameter}), suggesting that the information discarded by Eq.~\eqref{eq:x_cut} is small relative to that carried by the travel-time observations.

Conditional on $\boldsymbol{Z}$ and $\boldsymbol{\sigma}^2$, Eq.~\eqref{eq:x_cut} is the posterior of a linear Gaussian state-space model, so the entire trajectory $\{\boldsymbol{x}_t\}$ can be drawn from it exactly with the FFBS algorithm \citep{carter1994gibbs,fruhwirth1994data}.

For each time interval $t$, let $\boldsymbol{y}_t \in \mathbb{R}^{M_t}$ be the vector of all $M_t$ trip observations, $M_t=\sum_{(o,d)\in B}M_{odt}$. Conditional on the known path choices, the observation model is:
\begin{equation}
    p(\boldsymbol{y}_t \mid \boldsymbol{x}_t, \boldsymbol{Z}, \boldsymbol{\sigma}^2) = \mathcal{N}(\boldsymbol{A}_t\boldsymbol{x}_t, \boldsymbol{\Sigma}_t),
\end{equation}
where $\boldsymbol{A}_t$ is the routing matrix for all trips in interval $t$, $\boldsymbol{C}_t$ is the variance-design matrix whose $i$th row is the vector $\boldsymbol{c}_{od}^k$ associated with the current path assignment for trip $i$, and $\boldsymbol{\Sigma}_t=\text{diag}\{\boldsymbol{C}_t\boldsymbol{\sigma}^2\}$ is the diagonal observation covariance matrix. FFBS runs the Kalman filter forward over $t=1,\ldots,T$, initialized at $(\boldsymbol{m}_0,\boldsymbol{P}_0)$ and with prediction step $\boldsymbol{\mu}_{t\mid t-1}=\boldsymbol{\mu}_{t-1\mid t-1}$, $\boldsymbol{P}_{t\mid t-1}=\boldsymbol{P}_{t-1\mid t-1}+\text{diag}(\boldsymbol{\tau}^2)$, and then draws $\boldsymbol{x}_T,\ldots,\boldsymbol{x}_1$ backward from the smoothing distributions; the recursions are standard and are reproduced in Section~B of the Supplementary Material.

A direct implementation of this algorithm is computationally infeasible. The primary bottleneck is the inversion required to compute the Kalman gain $\boldsymbol{G}_t = \boldsymbol{P}_{t\mid t-1}\boldsymbol{A}_{t}^\top(\boldsymbol{A}_t\boldsymbol{P}_{t\mid t-1}\boldsymbol{A}_t^\top + \boldsymbol{\Sigma}_t)^{-1}$ (written $\boldsymbol{G}_t$ rather than $\boldsymbol{K}_t$ to avoid confusion with the kernel matrices). The matrix to be inverted has dimensions $M_t \times M_t$, where $M_t$ (the number of trips in an interval) can be on the order of hundreds of thousands. The $\mathcal{O}(M_t^3)$ complexity is prohibitive.

To overcome this, we leverage the information form of the Kalman filter, which reformulates the Kalman gain calculation using the Woodbury matrix identity. This shifts the computational burden from inverting an $M_t \times M_t$ matrix to inverting a $c \times c$ matrix, where $c$ is the number of network cost attributes:
\begin{equation}
    \boldsymbol{G}_t = (\boldsymbol{P}_{t\mid t-1}^{-1} + \boldsymbol{A}_t^\top\boldsymbol{\Sigma}_t^{-1}\boldsymbol{A}_t)^{-1}\boldsymbol{A}_t^\top\boldsymbol{\Sigma}_t^{-1}.
\end{equation}
Since $c$ is typically on the order of hundreds, the complexity is reduced to $\mathcal{O}(M_t c^2 + c^3)$, where the first term, for forming $\boldsymbol{A}_t^\top\boldsymbol{\Sigma}_t^{-1}\boldsymbol{A}_t$, is small in practice because $\boldsymbol{A}_t$ is sparse, making the FFBS algorithm efficient and scalable for large-scale transit networks.

\subsection{Sampling the Error Variances $\{\sigma_k^2\}$}
The full conditional posterior distributions for the error variances $\{\sigma_{ae}^2, \sigma_h^2, \sigma_u^2\}$ do not follow standard forms and therefore preclude direct sampling. While Metropolis-Hastings is a common alternative, it requires careful tuning of proposal distributions and step sizes to ensure efficient exploration of the parameter space.

To achieve a more robust and efficient sampler, we use slice sampling, a technique introduced by \cite{neal2003slice}. The key idea behind slice sampling is to sample from a complex distribution by introducing an auxiliary variable that transforms the problem into sampling uniformly from a region under the curve of the target distribution. This avoids the need for a carefully tuned proposal distribution as in Metropolis-Hastings, and it automatically adapts to the local shape of the posterior, often leading to better convergence properties. We apply slice sampling to update each error variance sequentially, conditional on the current values of all other parameters. The full conditional posterior for each variance parameter $\sigma_k^2$ is proportional to the product of its prior and the conditional likelihood of the observed data given all other parameters. For example, the full conditional posterior density of $\sigma_{ae}^2$ is:
\begin{equation}\label{eq:sigma_cond}
    p\left(\sigma_{ae}^2 \mid \cdot\right) \propto \prod_{t=1}^T p\left(\boldsymbol{y}_t \mid \boldsymbol{x}_t, \sigma_{ae}^2, \sigma_h^2, \sigma_u^2, \boldsymbol{Z}\right)\times p\left(\sigma_{ae}^2\right) = \prod_{t=1}^T \mathcal{N}\left(\boldsymbol{y}_t; \boldsymbol{A}_t\boldsymbol{x}_t, \text{diag}\{\boldsymbol{C}_t\boldsymbol{\sigma}^2\}\right)\times p\left(\sigma_{ae}^2\right).
\end{equation}
We use the shrinkage slice sampler of \citet{neal2003slice} with a randomly positioned initial bracket; the steps are listed as Algorithm~S1 in Section~C of the Supplementary Material.

A similar procedure is followed to sample from the full conditional posterior distributions of $\sigma_h^2$ and $\sigma_u^2$.

\subsection{Sampling the Tensor Parameters $\boldsymbol{U}, \boldsymbol{V}, \boldsymbol{W}, q_1, q_2$} \label{sec:sample_tensor}

The parameters governing the path choice model, the tensor factor matrices $\boldsymbol{U}, \boldsymbol{V}, \boldsymbol{W}$ and the baseline effects $q_1, q_2$, have non-conjugate full conditional posterior distributions. A key feature of our inference strategy is to integrate out the discrete latent path choices $\boldsymbol{Z}$ when sampling these parameters. This constitutes a \textit{collapsed Gibbs sampling} step, which can significantly improve the mixing and convergence rate of the sampler by reducing the conditioning between the path choices and the utility function parameters.

The primary challenge remains that the columns of the factor matrices $\boldsymbol{U}, \boldsymbol{V},$ and $\boldsymbol{W}$ are endowed with Gaussian priors that induce strong correlations. For such high-dimensional, correlated parameters, standard MCMC methods are inefficient. We therefore employ elliptical slice sampling \citep{murray2010elliptical}, an algorithm specifically designed for latent Gaussian models. It efficiently samples from posterior distributions where the prior is a multivariate normal. It generates proposals by defining an elliptical trajectory based on a random draw from the prior. More precisely, given a current state $\boldsymbol{v}_r$ and a Gaussian prior $\mathcal{N}\left(\boldsymbol{0}_n,\boldsymbol{K}_S\right)$, the sampler samples a direction $\boldsymbol{\nu}\sim \mathcal{N}\left(\boldsymbol{0}_n,\boldsymbol{K}_S\right)$, forming an ellipse through the linear combination $\boldsymbol{v}_r^\prime=\boldsymbol{v}_r \cos{\varphi}+\boldsymbol{\nu}\sin{\varphi}$, where $\varphi$ is an angle sampled from a slice defined by the posterior. The sampler iteratively shrinks the angular bracket until a point on the ellipse is accepted. This method automatically respects the correlation structure of the prior and requires no tuning, making it ideal for updating the columns of the factor matrices.

We update each factor matrix column-wise. For example, to sample column $\boldsymbol{v}_r$ from the spatial factor matrix $\boldsymbol{V}$, we sample from its full conditional posterior, which is proportional to the product of its GP prior and the marginal likelihood of the data:
\begin{equation}
    p(\boldsymbol{v}_r \mid \cdot) \propto \mathcal{N}(\boldsymbol{v}_r; \boldsymbol{0}_n, \boldsymbol{K}_S) \times L(\boldsymbol{v}_r),
\end{equation}
where the marginal likelihood $L(\boldsymbol{v}_r)$ is obtained by integrating out the path choices $\boldsymbol{Z}$. It should be noted that the choice model parameters are determined based on observations from O-D pairs with multiple feasible paths. Let $B_R=\left\{\left(o,d\right)\mid\left|R_{od}\right|>1,\left(o,d\right)\in B\right\}$ represent the set of O-D pairs that have more than one available path. The computation of likelihood $L\left({\boldsymbol{v}}_r\right)$ can be expressed as
\begin{align}
    L(\boldsymbol{v}_r \mid \cdot) &= p(\boldsymbol{Y} \mid \boldsymbol{U}, \boldsymbol{v}_r, \boldsymbol{V}_{:,-r}, \boldsymbol{W}, q_1, q_2, \{\boldsymbol{x}_t\}, \boldsymbol{\sigma}^2) \nonumber \\
    &= \prod_{\left(o,d\right)\in B_R}\prod_{t=1}^{T}\prod_{m=1}^{M_{odt}} p(y_{odt}^m \mid \cdot) \nonumber \\
    &= \prod_{\left(o,d\right)\in B_R}\prod_{t=1}^{T}\prod_{m=1}^{M_{odt}} \left[ \sum_{k=1}^{|R_{od}|} p(y_{odt}^m \mid z_{odt}^m=k, \cdot) \times p(z_{odt}^m=k \mid \boldsymbol{x}_t,\theta_{ot},\phi_{ot}) \right],\nonumber \\
    & r=1,2,\cdots, R.
\end{align}

This likelihood for each trip observation is a mixture of Gaussians, weighted by the MNL choice probabilities. The elliptical slice sampling procedure for each column of factor matrix $\boldsymbol{V}$ is that of \citet{murray2010elliptical} and is listed as Algorithm~S2 in Section~C of the Supplementary Material. The same procedure is used for the columns of $\boldsymbol{U}$ and $\boldsymbol{W}$, each with its respective Gaussian prior. The simpler scalar baseline parameters, $q_1$ and $q_2$, are updated using standard slice sampling.

\subsection{Sampling the Covariance Matrix $\boldsymbol{K}_U$}
The prior for the covariance matrix $\boldsymbol{K}_U$ is an inverse Wishart distribution, which is conjugate to the Gaussian prior on the factor vectors $\{\boldsymbol{u}_r\}$. Therefore, we can sample directly from the posterior of $\boldsymbol{K}_U$, which is also an inverse Wishart distribution with updated parameters:
\begin{equation}
    p(\boldsymbol{K}_U \mid \boldsymbol{U}) = \mathcal{W}^{-1}(\boldsymbol{\Omega}_0 + \boldsymbol{U}\boldsymbol{U}^\top, \nu_0 + R),
\end{equation}
where $(\boldsymbol{\Omega}_0, \nu_0)$ are the prior parameters and $R$ is the number of factors (the CP rank).

\subsection{Sampling the Path Choice Outcomes $\boldsymbol{Z}$}\label{sec:sample_z}
Conditional on all other parameters, the path choice $z_{odt}^m$ for each trip is a categorical variable that can be sampled independently. The posterior probability for a trip having taken a specific path $k$ is proportional to the product of two terms: (1) the likelihood of observing the travel time $y_{odt}^m$ given path $k$, and (2) the prior probability of choosing path $k$ from the MNL model.
\begin{equation}
    p(z_{odt}^m = k \mid \cdot) \propto p(y_{odt}^m \mid z_{odt}^m=k, \cdot) \times p(z_{odt}^m=k \mid \boldsymbol{x}_t,\theta_{ot},\phi_{ot}).
\end{equation}
The first term is the Gaussian density from Eq.~\eqref{eq:linear_Gaussian}, and the second is the MNL probability from Eq.~\eqref{eq:choice}. We compute this product for every feasible path $k \in R_{od}$ and sample $z_{odt}^m$ from the resulting normalized categorical distribution. This step is highly parallelizable across all trips.

\subsection{Overall Gibbs Sampling Algorithm}
The complete inference procedure is summarized in Algorithm~\ref{al:overall}. The sampler iterates through the blocks of parameters and latent variables as described in the preceding sections. After an initial burn-in period of $I_1$ iterations to ensure convergence, we collect additional $I_2$ samples to approximate the target posterior distribution.

\begin{algorithm}[!t]
\renewcommand{\algorithmicrequire}{\textbf{Input:}}
\renewcommand{\algorithmicensure}{\textbf{Output:}}
\caption{Overall modularized Gibbs sampling algorithm.}\label{al:overall}
\begin{algorithmic}[1]
\REQUIRE Observations $\boldsymbol{Y}$, prior specifications, burn-in iterations $I_1$, sampling iterations $I_2$.
\ENSURE Posterior samples $\{\boldsymbol{\Gamma}^{(i)}, \boldsymbol{Z}^{(i)}\}_{i=I_1+1}^{I_1+I_2}$.
\STATE Initialize all parameters $\boldsymbol{\Gamma}^{(0)}$ and latent path choices $\boldsymbol{Z}^{(0)}$.
\FOR{$i=1$ to $I_1 + I_2$}
    \STATE Sample time-varying costs $\{\boldsymbol{x}_t^{(i)}\}$ using the FFBS algorithm.
    \STATE Sample error variances $\{{\sigma_k^2}^{(i)}\}$ using slice sampling.
    \STATE Sample tensor factor matrices $\boldsymbol{U}^{(i)}, \boldsymbol{V}^{(i)}, \boldsymbol{W}^{(i)}$ using elliptical slice sampling.
    \STATE Sample baseline parameters $q_1^{(i)}, q_2^{(i)}$ using slice sampling.
    \STATE Sample covariance matrix $\boldsymbol{K}_U^{(i)}$ from its inverse Wishart posterior.
    \STATE Sample latent path choices $\boldsymbol{Z}^{(i)}$ from their categorical posterior.
    \IF{$i > I_1$}
        \STATE Store the current state of all parameters $\boldsymbol{\Gamma}^{(i)}$ and latent variables $\boldsymbol{Z}^{(i)}$.
    \ENDIF
\ENDFOR
\end{algorithmic}
\end{algorithm}

\section{Experiments}\label{sec:exp}

\subsection{Benchmark Models}
To quantify the benefits of modeling spatiotemporal heterogeneity, we compare our proposed framework against three simpler benchmarks. These models form a nested comparison, systematically re-introducing complexity to isolate the effects of spatial and temporal variation in passenger choice behavior.

All four models share the travel-time model, the random-walk prior and the inference scheme of Section~\ref{sec:inference}, and differ only in the coefficients of Eq.~\eqref{eq:utility}. \textbf{Model 1 (static, homogeneous)} sets $\theta_{ot}\equiv\theta$ and $\phi_{ot}\equiv\phi$ for all stations and intervals. \textbf{Model 2 (spatially varying)} sets $\theta_{ot}=\theta_o$ and $\phi_{ot}=\phi_o$, specific to the origin station but constant over the day. \textbf{Model 3 (temporally varying)} sets $\theta_{ot}=\theta_t$ and $\phi_{ot}=\phi_t$, common to all stations. \textbf{Model 4} is the proposed spatiotemporal model of Section~\ref{sec:model}.

\subsection{Simulation Study}\label{sec:simulation}

To evaluate the performance of our proposed framework, we conduct a comprehensive simulation study. The primary objective is to determine if our inference method can accurately recover the true latent network costs and spatiotemporally varying choice parameters under realistic, high-dimensional conditions. To ground our simulation in a practical setting, we use the network topology of the Hong Kong MTR system, one of the world's most complex and densely used urban rail networks (Supplementary Fig.~S2). The network comprises 90 stations, 10 bidirectional lines, 8,010 O-D pairs, and 15,754 unique paths. Our model considers 544 distinct network cost attributes (access, egress, in-vehicle, and transfer costs). We analyze a full day of operations, segmented into thirty-two 30-minute intervals from 6:00 AM to 10:00 PM ($T=32$).

We generate a synthetic dataset by first defining a set of ground-truth parameters. The time-varying network costs $\{\boldsymbol{x}_t\}$ are generated from a random walk process with noise $\boldsymbol{\tau}^2 = 25 \times \boldsymbol{1}_c$ (costs in seconds), starting from an initial state $\boldsymbol{m}_0$ derived from an analysis of the real-world data to ensure a reasonable baseline. The latent factor matrices ($\boldsymbol{U}, \boldsymbol{V}, \boldsymbol{W}$) are drawn from their respective priors (Gaussian, and GP with diffusion and squared-exponential kernels). The spatiotemporal path choice parameters ($\boldsymbol{\Theta}$ and $\boldsymbol{\Phi}$) are generated from the tensor factorization model with rank $R=4$ and average effects $q_1=-0.05$ and $q_2=-0.03$. Using these true parameters, we simulate a complete set of individual trip observations from the MNL choice model. This synthetic dataset then serves as the input for our inference algorithm. The same fixed values of $\boldsymbol{\tau}^2$ and $(\boldsymbol{m}_0,\boldsymbol{P}_0)$ are used in the inference.

{We apply our proposed Bayesian inference method to estimate the latent costs and choice parameters. To assess mixing rather than relying on a single trajectory, we run two independent chains, each for 10,000 iterations, discard the first 8,000 iterations as burn-in, and retain the final 2,000 draws for posterior summaries. Section~A of the Supplementary Material reports trace and autocorrelation plots for the three variance components. We also report both split and rank-normalized Gelman-Rubin diagnostics \citep{vehtari2021rank}. The resulting split (rank-normalized) $\hat R$ values are 1.001 (1.001) for $\sigma_h^2$, 1.001 (1.001) for $\sigma^2_u$, and 1.000 (1.000) for $\sigma^2_{ae}$; the corresponding effective sample sizes are 525.9, 694.1, and 813.3. These values are close to one and are accompanied by effective sample sizes well above 200, providing empirical evidence of good mixing for the reported variance parameters \citep{vehtari2021rank}.  These results indicate that the posterior samples provide a reliable approximation of the underlying distributions for these parameters. We then compare the posterior estimates against the known ground-truth values. }

\begin{figure*}[!t]
\centering
\includegraphics[width=0.95\textwidth]{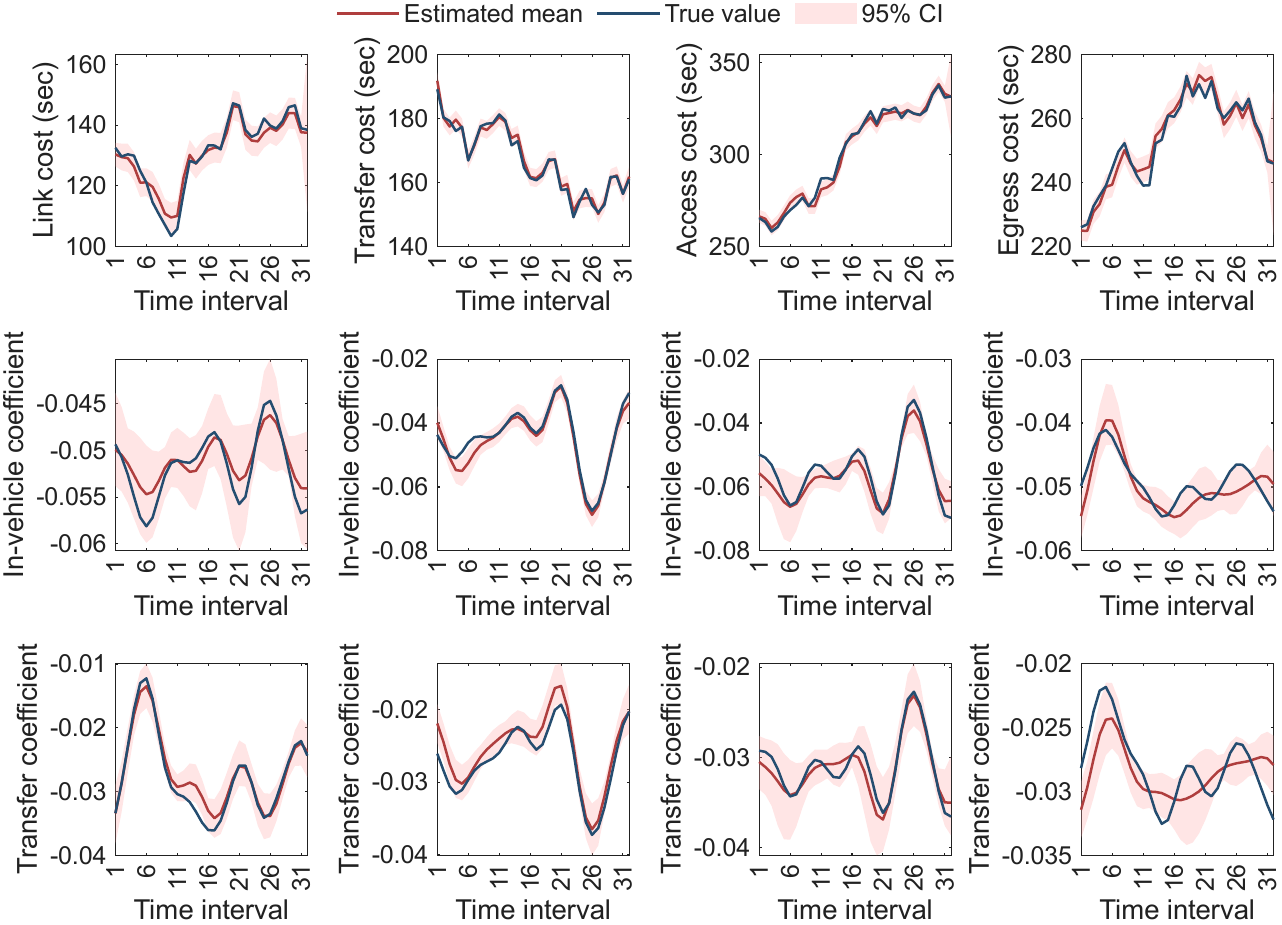}
\caption{Recovery of simulated parameters. The posterior means (red lines) closely track the true values (blue lines) for both network costs and choice coefficients, with the 95\% credible intervals (shaded areas) providing good coverage.}
\label{fig:sim_parameter}
\end{figure*}

\begin{figure*}[!h]
\centering
\includegraphics[width=0.98\textwidth]{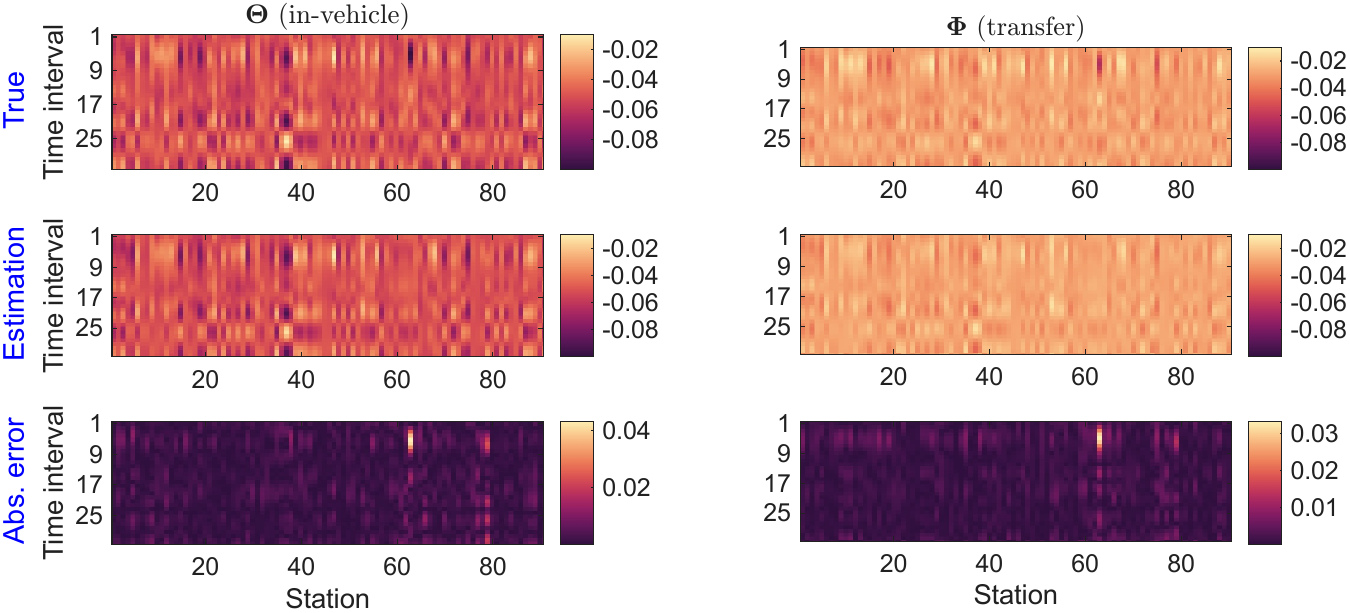}
\caption{Recovery of spatiotemporal choice coefficients. Top row: True values for in-vehicle ($\boldsymbol{\Theta}$) and transfer ($\boldsymbol{\Phi}$) coefficients. Middle row: Posterior mean estimates. Bottom row: Absolute estimation errors, which are generally low across space and time.}
\label{fig:est_error}
\end{figure*}

The results demonstrate a successful recovery of the true parameters. Fig.~\ref{fig:sim_parameter} shows the posterior means and 95\% credible intervals for the network costs and the choice coefficients at four representative stations over time. The estimated means closely track the true temporal trends, and the credible intervals consistently cover the true values, suggesting reasonable uncertainty quantification in the simulation. Fig.~\ref{fig:est_error} provides a more granular assessment of the spatiotemporal choice coefficients ($\boldsymbol{\Theta}$ and $\boldsymbol{\Phi}$). The heatmaps of the absolute estimation error reveal generally low errors across all stations and time intervals, confirming the model's ability to recover fine-grained behavioral heterogeneity.

\subsection{Application to Real-world Metro Data}

To test our model's predictive capabilities, we apply it to a large-scale dataset from the Hong Kong MTR system. The dataset comprises over four million AFC records from a typical weekday (May 10, 2018). The records were provided by MTR Corporation under a confidentiality agreement and cannot be redistributed. Our evaluation focuses on a key task: predicting the travel time distribution for out-of-sample trips.

To construct a robust validation experiment, we first partition the data into training (90\%) and validation (10\%) sets. This partitioning is performed via stratified sampling on the set of O-D pairs that have more than one feasible path. This strategy ensures that the validation set contains a representative sample of trips where path choice is a non-trivial component, allowing for a rigorous test of the model's ability to capture complex travel behavior.

After fitting the model on the training data, we evaluate its performance on the validation set. For each trip in the validation set, our model produces a full posterior predictive distribution for its travel time. This distribution is a mixture model that correctly accounts for uncertainty from three sources: (1) the unobserved path choice, (2) the observation error, and (3) the posterior uncertainty in all model parameters ($\boldsymbol{\Theta}, \boldsymbol{\Phi}, \{\boldsymbol{x}_t\}, \boldsymbol{\sigma}^2$). Formally, the distribution for a new observation $y_{odt}^{\text{new}}$ is given by:
\begin{equation}
    p(y_{odt}^{\text{new}} \mid \boldsymbol{Y}_{\text{train}}) = \int \left[ \sum_{k=1}^{|R_{od}|} p(z_{odt}^{\text{new}}=k \mid \boldsymbol{x}_t,\theta_{ot},\phi_{ot})\, \mathcal{N}(y_{odt}^{\text{new}}; \boldsymbol{A}_{od}^k\boldsymbol{x}_t, \boldsymbol{c}_{od}^k\boldsymbol{\sigma}^2) \right] p(\boldsymbol{\Gamma} \mid \boldsymbol{Y}_{\text{train}})\, d\boldsymbol{\Gamma}, \label{eq:mixture}
\end{equation}%
where $\boldsymbol{\Theta}$ and $\boldsymbol{\Phi}$ are computed from $\boldsymbol{\Gamma}$ through Eq.~\eqref{eq:tensor-1}. Although this integral does not have a closed form, once samples from the posterior distribution of $\boldsymbol{\Gamma}$ are available, samples from the posterior predictive distribution are easily obtained using compositional sampling.

We assess the quality of these predictions using three standard metrics. To evaluate point-prediction accuracy, we use the Root Mean Squared Error (RMSE) and Mean Absolute Error (MAE) in the validation set and the mean of the posterior predictive distribution ($\hat{y}_{odt} = E[y_{odt}^{\text{new}}\mid \boldsymbol{Y}_{\text{train}}]$).

To evaluate the accuracy of the full probabilistic prediction, we use the Continuous Ranked Probability Score (CRPS) \citep{gneiting2007strictly}. The CRPS is a proper scoring rule that generalizes the MAE to probabilistic predictions, measuring the discrepancy between the predictive cumulative distribution function (CDF), $F_{odt}$, and the empirical CDF of the observation. For a single observation $y_{odt}^m$, it is defined as:
\begin{equation}
    \text{CRPS}(F_{odt}, y_{odt}^m) = \int_{-\infty}^{\infty} (F_{odt}({s}) - \mathbb{I}({s} \geq y_{odt}^m))^2 d{s},
\end{equation}
where $\mathbb{I}(\cdot)$ is the indicator function. We report the average CRPS over all observations in the validation set. Lower values for all three metrics indicate better predictive performance.

\subsection{Model Performance Evaluation}

The results of our model comparison, presented in Table~\ref{tab:performance_five}, confirm the critical importance of modeling spatiotemporal heterogeneity. Our full proposed model (Model 4) consistently outperforms all simpler benchmarks across all evaluation metrics (MAE, RMSE, and CRPS) and for all tested tensor ranks ($R=1,2,4,6$). Both the spatially varying model (Model 2) and the temporally varying model (Model 3) outperform the static model (Model 1), with Model 2 achieving slightly better results than Model 3. The proposed spatiotemporal model (Model 4) achieves the best performance across all three metrics. This clear progression underscores that both spatial and temporal dynamics are essential, and that accurately capturing their interaction is key to achieving the highest predictive accuracy. The results also show that while performance generally improves with a higher tensor rank $R$, the gains diminish, allowing for a principled trade-off between model complexity and performance.

\begin{table}[!h]
\caption{The performance of trip travel time prediction (in seconds) with various models.}
\label{tab:performance_five}
\small
\centering
\begin{tabular}{cccccccc}
\toprule
\multirow{2}{*}{} & \multirow{2}{*}{Model 1} & \multirow{2}{*}{Model 2} & \multirow{2}{*}{Model 3} & \multicolumn{4}{c}{Model 4 (proposed model)}
\\
\cmidrule{5-8}
\multicolumn{1}{c}{}&{}&{}&{}&$R=1$&$R=2$&$R=4$&$R=6$\\
\midrule
MAE & 145.92 & 138.54 & 139.65 & 126.28 &  125.56 & 124.76 & \textbf{124.24} \\
RMSE & 234.95 & 206.54 & 213.27 & 169.48 &  168.24 & 166.64 &\textbf{165.18} \\
CRPS & 108.53 & 101.82 & 103.15 & 91.46 &  90.61 & 90.03 & \textbf{89.64} \\
\bottomrule
\multicolumn{8}{l}{Lowest scores in bold.}
\end{tabular}
\end{table}

{

CRPS improves alongside the point metrics, indicating the gains extend to the predictive distribution, not only its mean. This is particularly relevant in the present setting, because the main goal is not simply point prediction of trip travel time, but inference on latent passenger path choices and dynamic network costs, for which trip travel time serves only as an indirect validation target.  Therefore, the practical value of the proposed framework lies not only in average predictive performance, but also in its ability to provide behaviorally realistic passenger assignment, interpretable variation in path choice sensitivities, and uncertainty quantification for latent network conditions and route choices.}

For the best-performing specification (Model 4, $R=6$), the posterior predictive means track the observed travel times closely ($R^2=0.95$; Supplementary Fig.~S3), and the prediction errors grow with trip length, as implied by the heteroscedastic variance model of Section~\ref{sec:model}.

{To further assess the reliability of the proposed framework for passenger trip assignment, we show in Supplementary Fig. S4 the spatial distribution of station-level travel time estimation errors at three representative time periods. The results show that errors are generally small across the network, without a systematic concentration of large errors at major transfer hubs. This is particularly encouraging for passenger trip assignment, as these stations often connect multiple competing paths and play a critical role in distinguishing passengers' route choices. Despite the low-rank constraint, the model maintains comparable estimation accuracy across stations with different network characteristics, suggesting that the low-rank representation preserves important spatiotemporal information relevant to path inference. Although some localized variations remain, their spatial patterns do not indicate a consistent deterioration at complex interchange stations. Together with the overall model performance, these findings provide supporting evidence for the robustness of the proposed framework in inferring passenger path choices across a complex metro network.
}

\subsection{Interpreting Model Parameters}

Beyond estimation, our model provides rich, interpretable insights into passenger behavior and network dynamics. We analyze the model parameters estimated from the real-world data.

\begin{figure*}[!t]
\centering
\includegraphics[width=0.95\textwidth]{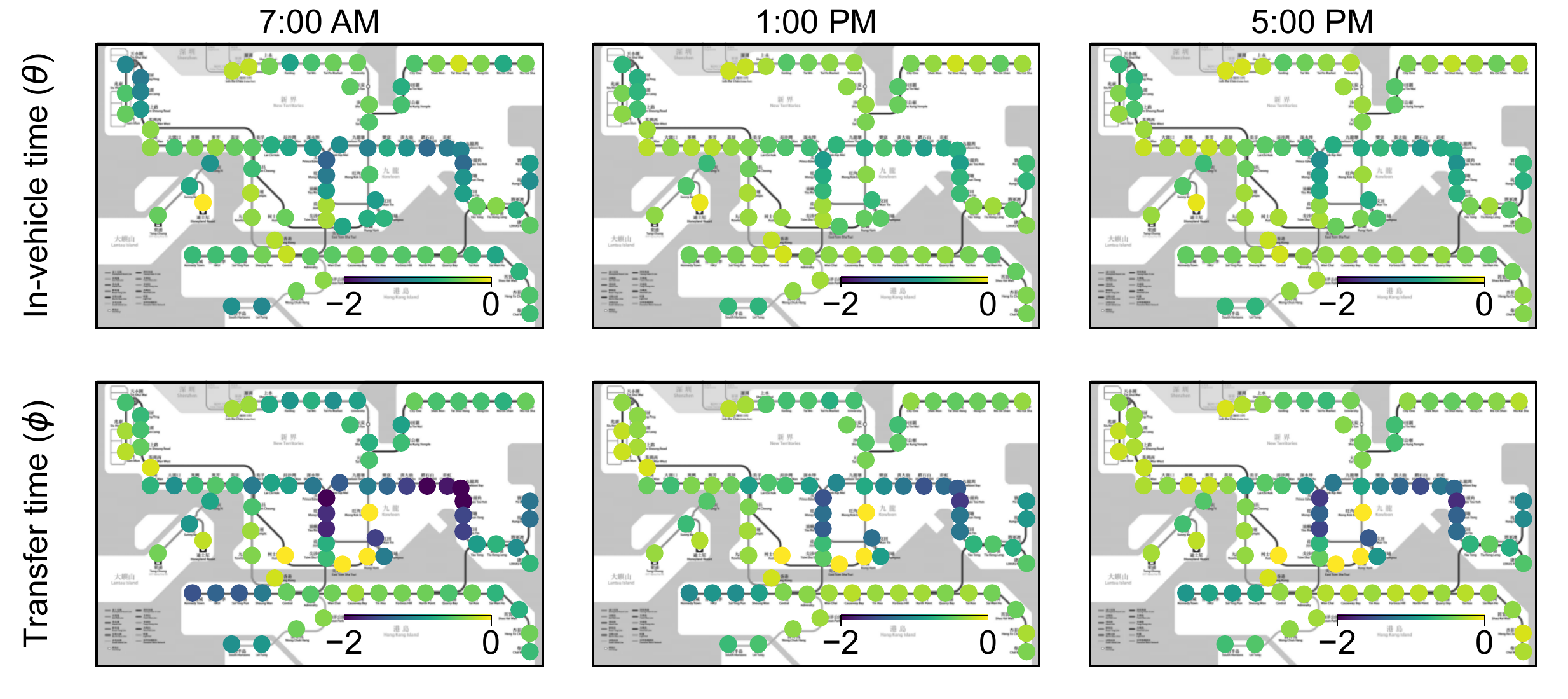}
\caption{The estimated posterior mean of spatiotemporal $\boldsymbol{\Theta}$ (in-vehicle time) and $\boldsymbol{\Phi}$ (transfer time) in the choice model. There are two rows: the first row represents the posterior mean of the coefficients associated with in-vehicle costs; the second row represents the posterior mean of the coefficients associated with transfer costs. The first column presents coefficients for the interval between 7:00 AM and 7:30 AM during morning peak hours; the second column presents coefficients for the interval between 1:00 PM and 1:30 PM during off-peak hours; the last column shows coefficients for the interval between 5:00 PM and 5:30 PM during afternoon peak hours.}
\label{fig:sp_invehicle}
\end{figure*}

The estimated coefficients for in-vehicle time ($\boldsymbol{\Theta}$) and transfer time ($\boldsymbol{\Phi}$) reveal pronounced spatiotemporal heterogeneity in passenger preferences (see Supplementary Figs.~S6-S8). It should be noted that the costs in this experiment are measured in minutes, so the coefficients are per minute. As shown in Fig.~\ref{fig:sp_invehicle}, these preferences evolve distinctly over the course of the day. During the morning peak (7:00 AM), the coefficients become more negative, indicating a heightened sensitivity to travel time. This reflects passengers' desire for efficiency during commute hours. The aversion to transfers ($\boldsymbol{\Phi}$) is particularly strong during peak periods, when network congestion makes transfers most exhausting. In contrast, during off-peak hours (1:00 PM), passengers are more tolerant of longer travel and transfer times.

Spatially, the model uncovers a clear distinction in preferences based on trip origin. For example, sensitivity to in-vehicle time is lower (less negative) for trips originating at peripheral stations compared to those from centrally located stations. This is intuitive: for longer trips that are typical from the periphery, extended in-vehicle durations are an expected and unavoidable component. Conversely, for the shorter trips common from central areas, any additional in-vehicle time can feel disproportionately inconvenient. A similar pattern holds for transfer sensitivity, which is higher (more negative) at central stations. Passengers starting centrally, who expect shorter and more direct journeys, penalize transfers more heavily than those on longer, multi-stage trips from the network's edge. This alignment between in-vehicle and transfer time sensitivities reveals the model's ability to capture the interconnected and location-dependent priorities of passengers.

\begin{figure*}[!t]
\centering
\includegraphics[width=0.9\textwidth]{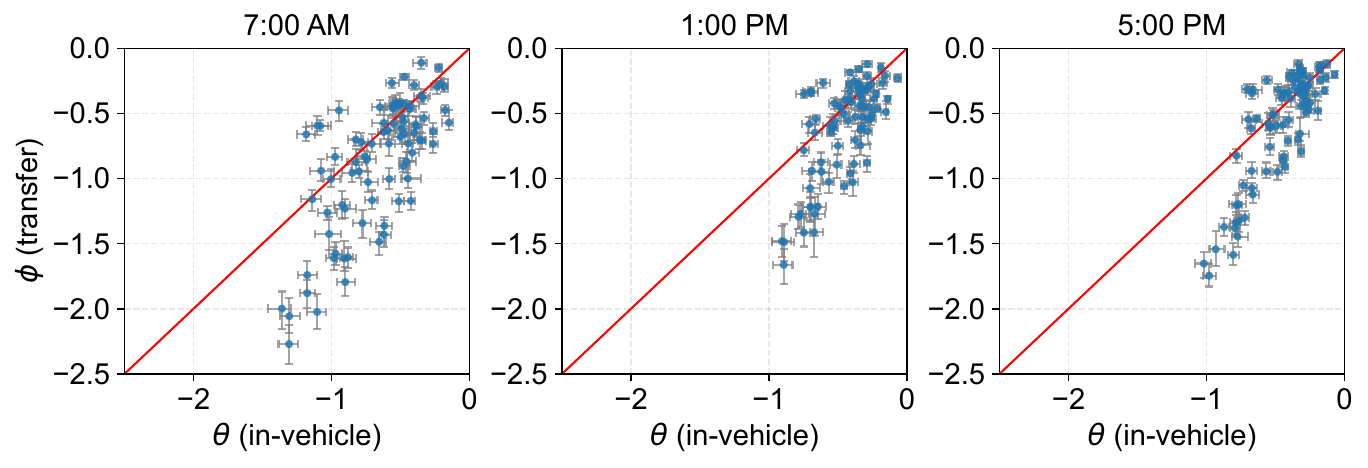}
\caption{Scatterplots of the estimated posterior means of spatiotemporal coefficients $\boldsymbol{\Theta}$ (in-vehicle time) and $\boldsymbol{\Phi}$ (transfer time) for each station, along with their associated 95\% credible intervals. Each subplot corresponds to a different time period: 7:00-7:30 AM (morning peak), 1:00-1:30 PM (off-peak), and 5:00-5:30 PM (afternoon peak), from left to right. Each point represents a station, with horizontal and vertical error bars indicating the 95\% credible intervals of $\theta$ and $\phi$, respectively. The red diagonal line denotes the identity line $\theta = \phi$ to guide comparison between the two coefficients.}
\label{fig:theta_phi_scatter}
\end{figure*}

While Fig.~\ref{fig:sp_invehicle} illustrates the estimated spatiotemporal coefficients separately, Fig.~\ref{fig:theta_phi_scatter} provides a direct comparison between the estimated in-vehicle coefficients ($\boldsymbol{\Theta}$) and transfer coefficients ($\boldsymbol{\Phi}$) at the station level. Across all time periods, the scatterplots reveal that the majority of points fall below the identity line, confirming that passengers generally perceive transfer time as less desirable than in-vehicle time. 

\begin{figure*}[!h]
\centering
\includegraphics[width=0.9\textwidth]{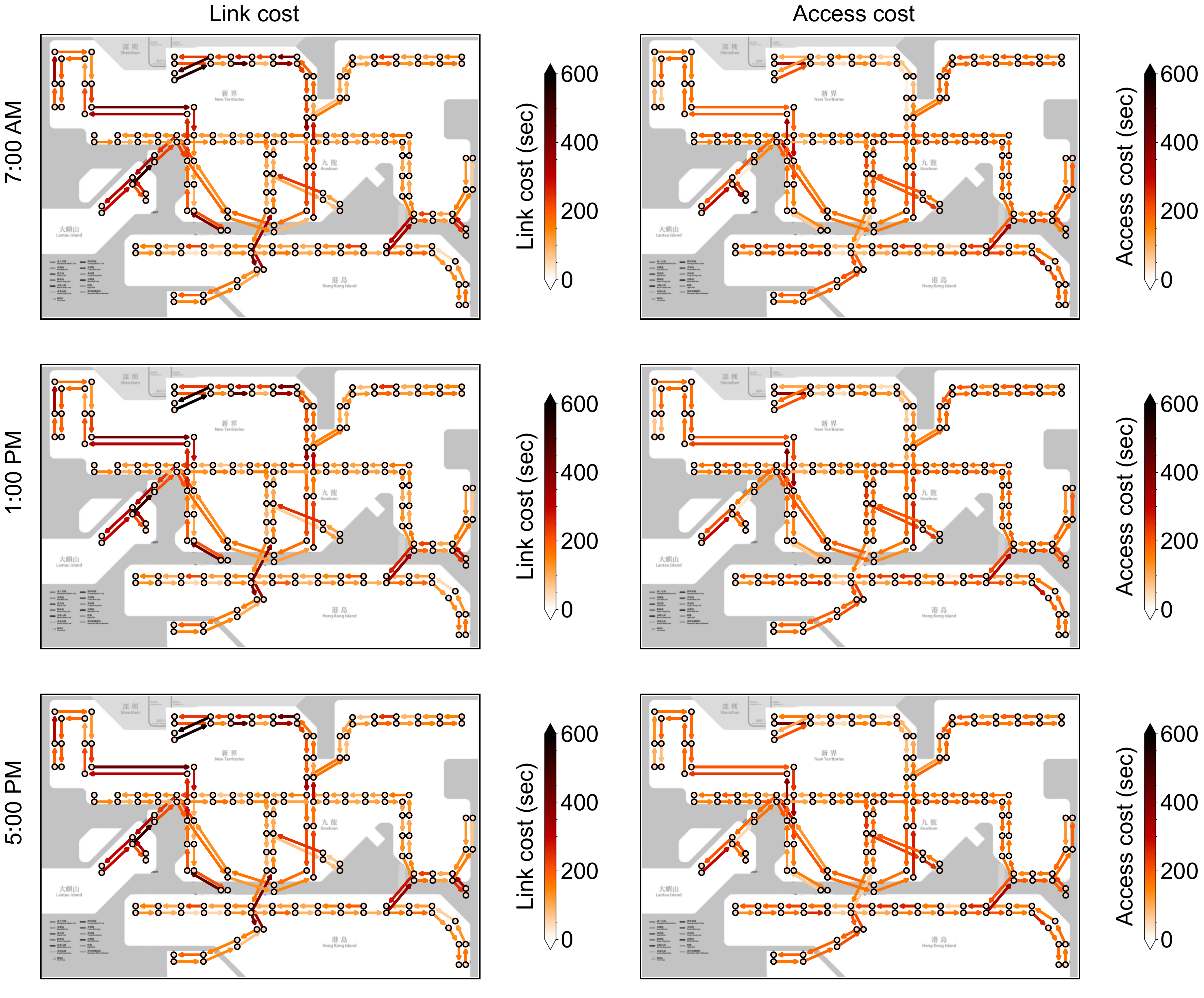}
\caption{Estimated link and access costs in the metro network.}
\label{fig:link_cost}
\end{figure*}

Our framework also provides a detailed view of the network's operational dynamics. The estimated network costs vary substantially throughout the day, reflecting the influence of demand patterns and operational adjustments (Fig.~\ref{fig:link_cost} and Supplementary Fig.~S5). As expected, many costs increase during peak hours due to higher passenger volumes and congestion. Interestingly, the model also captures instances where costs, particularly for access, decrease during peak periods, which may reflect more frequent service or more coordinated train arrivals that reduce waiting times.

Furthermore, the model quantifies the residual variability associated with passenger travel through the variances: the posterior means and 95\% credible intervals are $\sigma_h^2=0.0785$ $[0.0774,\,0.0797]$, $\sigma_u^2=1.4221$ $[1.4124,\,1.4321]$, and $\sigma_{ae}^2=1.6774$ $[1.6720,\,1.6832]$ min$^2$. In-vehicle components exhibit the smallest residual variability, highlighting the reliability and control of train movements. Transfer components show moderate residual variability, likely driven by factors like platform congestion and the synchronization of connecting services. The combined access-egress component exhibits the largest residual variability, which is reasonable given that entry and exit times depend on heterogeneous passenger walking speeds and the complexity of station layouts. Together, these findings demonstrate that our model not only estimates the dynamic costs but also provides a meaningful representation of the distinct sources of uncertainty inherent in passenger travel.

\subsection{Passenger Trip Assignment}

\begin{figure*}[!t]
\centering
    \centering
    \includegraphics[width=0.9 
    \textwidth]{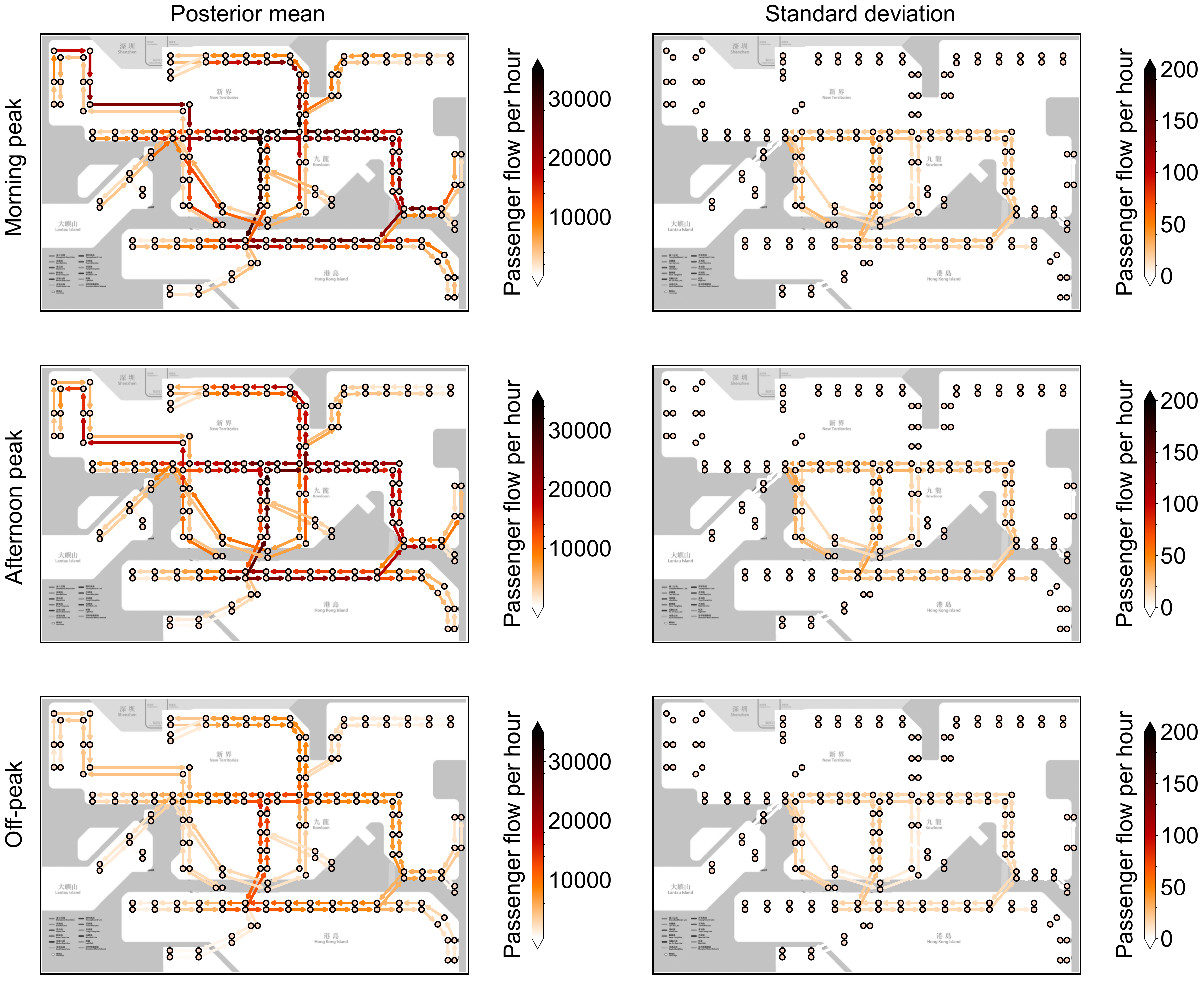}
\caption{The posterior means (first column) and pointwise standard deviations (second column) of passenger flow assignment for different time intervals.}
\label{fig:inbound_MP_flow_assignment}
\end{figure*}

The ultimate output of our model is a complete, probabilistic assignment of passenger trips to paths across the network. The panels of Fig.~\ref{fig:inbound_MP_flow_assignment} provide the posterior mean and standard deviation of these flow assignments, revealing clear spatiotemporal patterns. The mean flows align with typical daily commuting behavior: during the morning peak, inbound corridors leading to central business districts (e.g., Admiralty and Central) experience the highest traffic; this pattern reverses during the afternoon peak as passengers return to residential areas. Off-peak hours show lower and more balanced flows. The corresponding standard deviation plots provide a valuable quantification of uncertainty, highlighting that variability is highest at key transfer stations and along major corridors, particularly during congested peak hours.

To underscore the value of incorporating observed data, we compare the posterior summaries with the prior means and standard deviations of link flows shown in Supplementary Fig. S9. The prior assignment results are obtained by allocating passenger trips to paths based on the spatiotemporal MNL evaluated at the posterior draws of $\boldsymbol{\Theta}$, $\boldsymbol{\Phi}$ and $\{\boldsymbol{x}_t\}$, without using the travel time likelihood in Eq.~\eqref{eq:linear_Gaussian} to update the path choices. While the prior means capture  general commuting patterns, they are associated with considerably higher standard deviations, reflecting substantial uncertainty in the absence of travel time data. In contrast, the posterior standard deviations are noticeably smaller across the network, especially along major corridors. This reduction in uncertainty demonstrates the richness of the observed data and highlights the capability of our method to effectively leverage this information to improve the accuracy of passenger path choice estimates, thereby enhancing confidence in the resulting flow assignments.

\begin{figure*}[!t]
    \centering
 \includegraphics[width=0.8
    \textwidth]{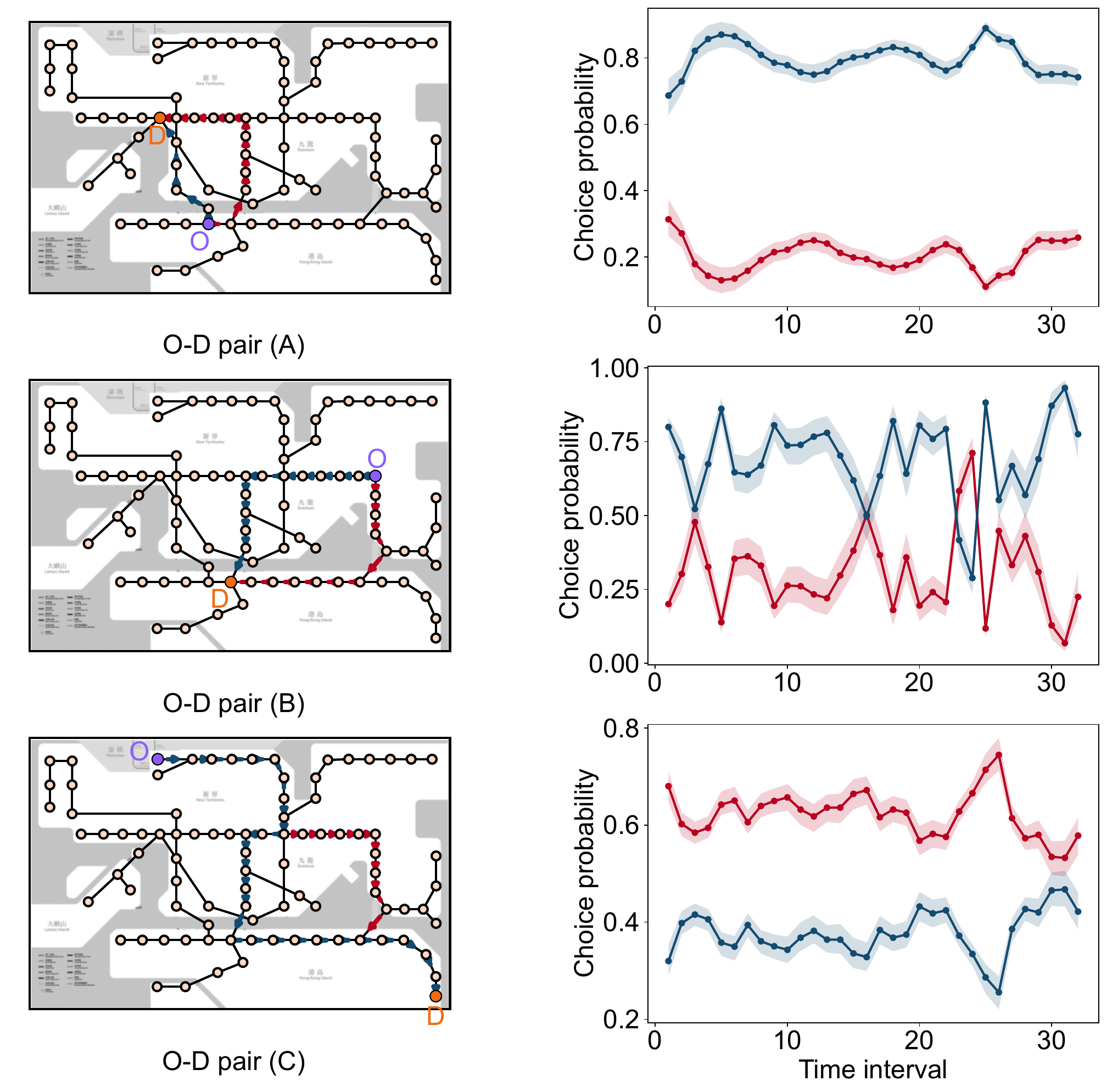}
\caption{Temporal path choice probabilities for three O-D pairs. Left panels show the paths on the metro network; right panels show their time-varying choice probabilities, with 95\% credible bands.}
\label{fig:ODpairp}
\end{figure*}

Fig.~\ref{fig:ODpairp} illustrates the temporal evolution of path choice probabilities for three selected O-D pairs. These plots show pronounced temporal heterogeneity, driven by the dynamic interplay between network costs and passenger preferences. For O-D pair (A), the dark-blue path is preferred throughout the day, most strongly during the morning and afternoon peaks. For pair (B), the choice probabilities for two competing paths cross at different times of the day, indicating a switch in preference as network conditions change. For pair (C), passengers consistently prefer the red path, especially during the afternoon peak. These results underscore the necessity of a dynamic modeling approach, as they reveal how passenger path choices fluctuate in response to real-time network conditions.

\section{Conclusion and Discussion}\label{sec:con}

This paper develops a novel Bayesian framework that addresses a fundamental challenge in transit modeling: the joint estimation of dynamic network costs and spatiotemporal passenger path choices from AFC data. Our approach addresses the key limitations of prior work by explicitly modeling the dynamic nature of both network conditions and passenger preferences. By representing trip travel times as a combination of dynamic cost components and parameterizing a multinomial logit choice model with spatiotemporally varying coefficients, we provide a realistic and comprehensive view of passenger behavior. The use of kernelized tensor factorization with GP priors makes this high-dimensional problem computationally tractable, while a tailored MCMC algorithm ensures efficient inference.

We demonstrated the model's superior predictive performance through a comprehensive evaluation on real-world data from the Hong Kong MTR system, where it substantially outperformed simpler, less dynamic benchmarks. Our results confirm that simultaneously capturing spatiotemporal variations in both network costs and passenger behavior is essential for achieving accurate flow assignments. The analysis of the model's outputs provides valuable insights into the dynamics of the metro system, from the evolution of passenger sensitivities to travel time, to the identification of key corridors and periods of network variability.

The practical implications of this framework for transit agencies and policymakers are substantial. First, the ability to produce accurate, dynamic passenger flow assignments provides a robust tool for optimizing operations. By understanding precisely how and when passengers travel, agencies can improve train scheduling, enhance transfer coordination, and manage service frequency to better match demand. Second, the model offers actionable insights into network performance by identifying bottlenecks, quantifying the variability of different travel components, and revealing how passenger preferences change across the network. These insights can inform targeted interventions, from station redesigns to congestion management strategies.
Third, because the choice coefficients are smooth over the network, the model could be used to extrapolate path choices to proposed new stations or lines, supporting service design and capacity planning. Finally, in multi-operator transit systems, the ability to accurately assign flows to specific network segments is crucial for supporting fair and transparent revenue sharing and benefit allocation agreements \citep{liu2024urban}, fostering better collaboration among stakeholders.

While our framework demonstrates strong capabilities, it also opens several avenues for future research. {An important limitation of the current framework is its computational cost. The most challenging component of the inference is the estimation of the high-dimensional dynamic network costs $\left\{\boldsymbol{x}_t\right\}_{t=1}^T$, since these latent states must be updated jointly across time and are directly linked to a very large number of trip-level observations. Although the proposed FFBS algorithm, implemented in information form with the Woodbury identity, substantially reduces the computational burden by replacing the inversion of an $M_t \times M_t$ matrix with that of a $c \times c$ matrix, posterior sampling remains computationally intensive for large-scale metro systems. In this sense, the current Bayesian assignment framework is an offline analysis tool rather than a real-time assignment tool. In many practical settings, transit agencies use passenger assignment results for downstream tasks such as service planning, operational diagnosis, bottleneck identification, and policy evaluation, where immediate real-time outputs are not required. For such applications, the additional computation is often acceptable because the Bayesian approach provides a full posterior distribution of the latent network costs, path choices, and behavioral parameters, thereby enabling uncertainty quantification in addition to point estimation. Future work could further improve computational scalability through parallel implementation, variational approximations, sparse or reduced-rank Gaussian process representations, and other approximate Bayesian inference strategies. A related direction concerns the modularized update of the network costs in Section~\ref{sec:sample_x}: a semi-modular scheme that tempers, rather than removes, the feedback from the choice model \citep{carmona2020semi}, or an exact Metropolis-Hastings correction applied one interval at a time, would make it possible to quantify the effect of the cut directly.}

A second limitation of the current model is the assumption that each cost type shares an error variance parameter. This parsimonious specification was adopted to keep inference feasible in a large-scale network, while still allowing travel-time variance to increase with path complexity. The simplification applies to the residual variance model, not to the mean network structure: spatial and temporal bottlenecks are primarily captured through the latent dynamic costs. Nevertheless, this assumption may smooth residual variance differences across network segments, such as between older and newer lines. A useful extension for future work would therefore be to allow subgroup-specific or hierarchically structured error variance parameters, such as line-specific, station-cluster-specific, or time-of-day-specific parameters, when sufficient data are available.

{Third, the model is designed for regular operating days. The Gaussian random-walk prior favors gradual temporal changes and may not adequately capture abrupt cost jumps caused by line failures, severe weather, or sudden crowding. We regard disruption-aware inference as outside the scope of the present study. A natural extension is a regime-switching state model in which the transition variance and drift depend on an observed or latent operating regime; jump-diffusion or heavy-tailed innovations could provide an alternative representation of sudden shocks. These extensions, combined with real-time data sources such as train tracking and incident reports, would allow route probabilities and their uncertainty to react quickly to disruptions and would permit a direct evaluation of how spatially clustered residual error affects routing reliability under abnormal conditions.}

Finally, our model currently assumes a general set of passenger preferences. Future work could relax this assumption by incorporating observed or latent heterogeneity across different passenger groups based on demographics, trip purpose, or fare class. Introducing this level of detail would allow for a more nuanced understanding of travel behavior and further improve the model's explanatory power and interpretability.

\section*{Acknowledgements}
This research is supported in part by the Canadian Statistical Sciences Institute (CANSSI) Collaborative Research Teams (CRT) grants and in part by the Natural Sciences and Engineering Research Council (NSERC) of Canada. 

\bibliography{reference}



\section*{Supplementary material}

# Supplementary appendices to Bayesian spatiotemporal modeling of passenger trip assignment in metro networks

It provides additional diagnostics, estimation results, and visualization of the posterior inference that support the findings in the main text.

## Appendix A. Markov chains and effective sample size

This section provides convergence diagnostics of the MCMC algorithm used for posterior inference:

- Fig. A.1 shows trace plots (Markov chains) for selected spatiotemporal parameters from two independent MCMC chains, demonstrating convergence and mixing.

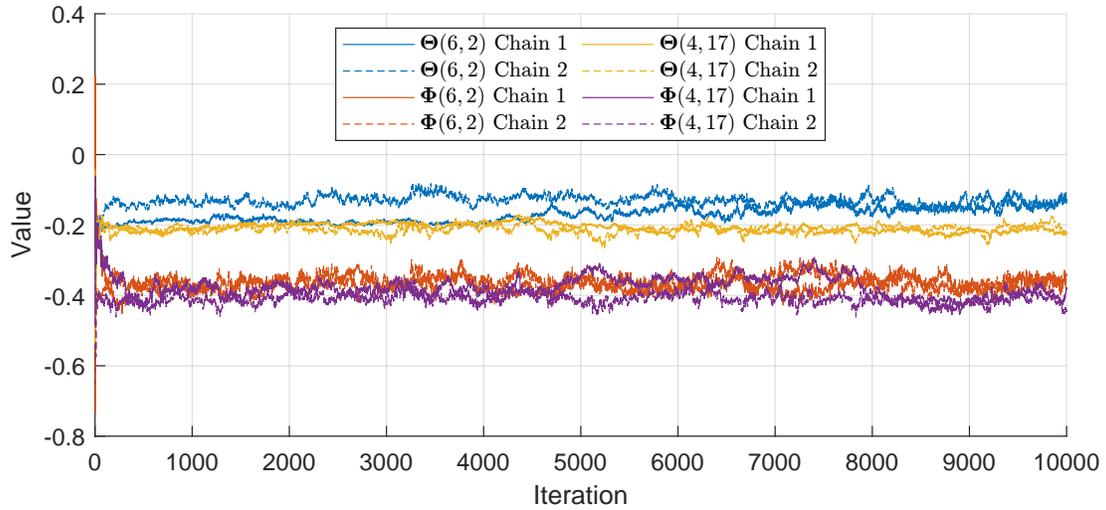

Fig. A.1: Markov chains for parameters.

- Figs. A.2 and A.3 and display heatmaps of the effective sample size for choice model parameters $\boldsymbol{\Theta}$ and $\boldsymbol{\Phi}$, as well as the time-varying network cost parameters $\boldsymbol{x}_t$, across all stations and time intervals. High ESS values indicate good mixing.

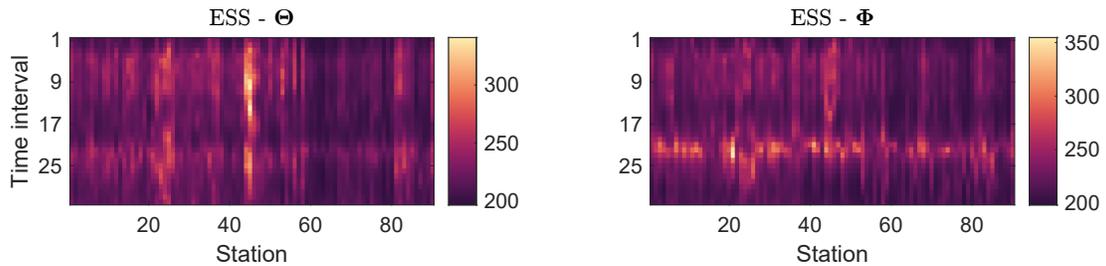

Fig. A.2: Effective sample size for choice parameters.

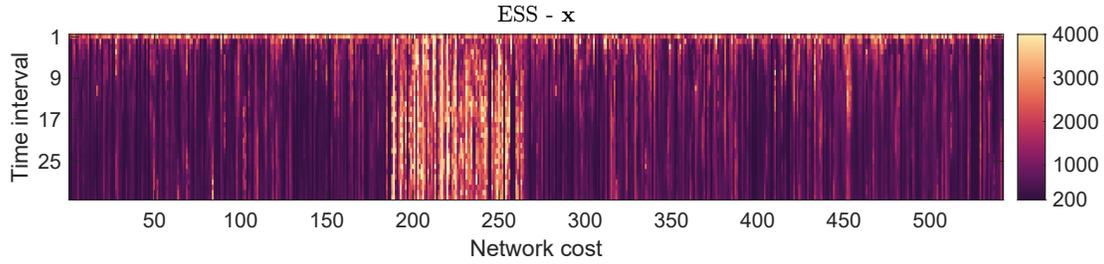

Fig. A.3: Effective sample size for network cost parameters.

## Appendix B. Estimated time-varying network costs and spatiotemporal coefficients in the choice model

This section provides visualizations of the posterior estimates of key model components. These posterior summaries confirm the model's ability to capture complex spatiotemporal variation in both passenger choice behavior and travel cost dynamics.

- Figs. B.1–B.3 show the posterior mean and 95% credible intervals (lower and upper bounds) for the spatiotemporal choice model parameters $\boldsymbol{\Theta}$ (in-vehicle cost sensitivity) and $\boldsymbol{\Phi}$ (transfer sensitivity) across all stations and time periods.

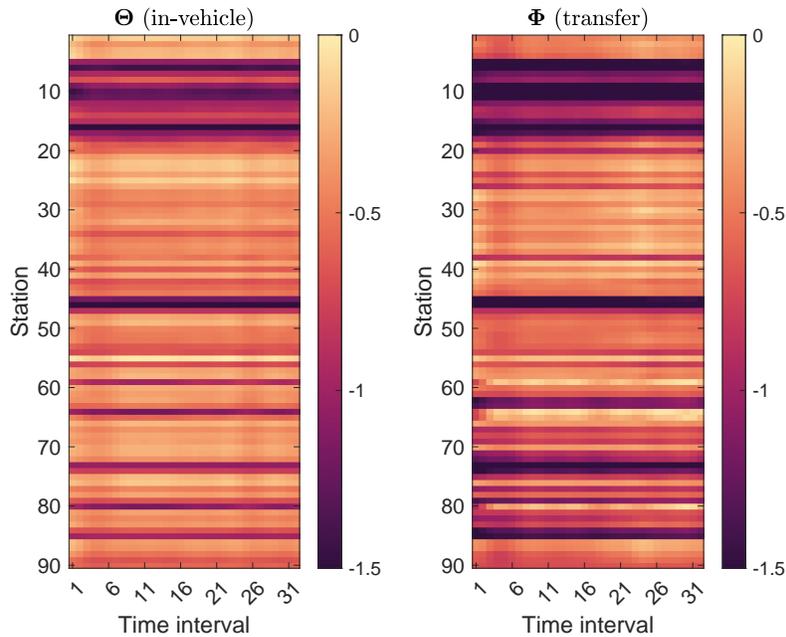

Fig. B.1: The estimated posterior mean of spatiotemporal parameters in the choice model.

- Fig. B.4 presents the posterior mean trajectories of the time-varying network costs, including link, transfer, access, and egress components.

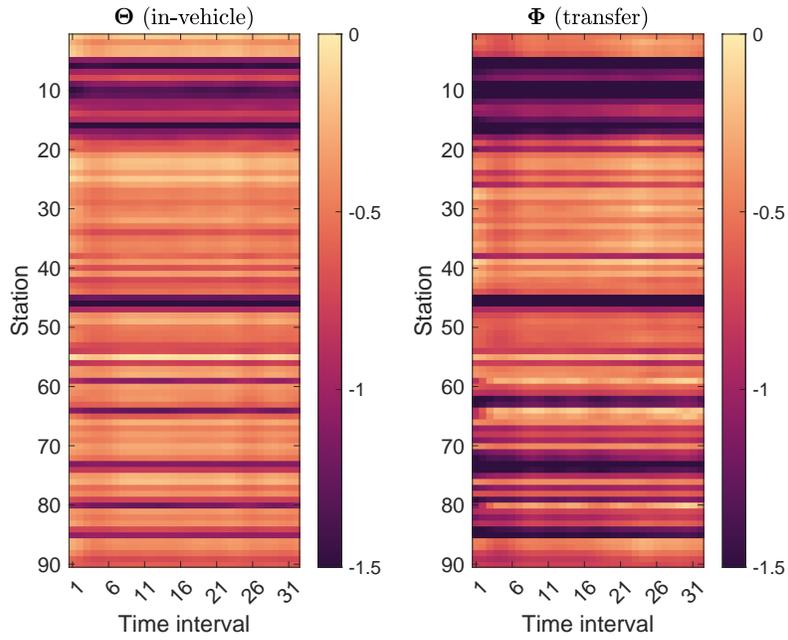

Fig. B.2: The lower bound of 95% credible interval of spatiotemporal choice parameters.

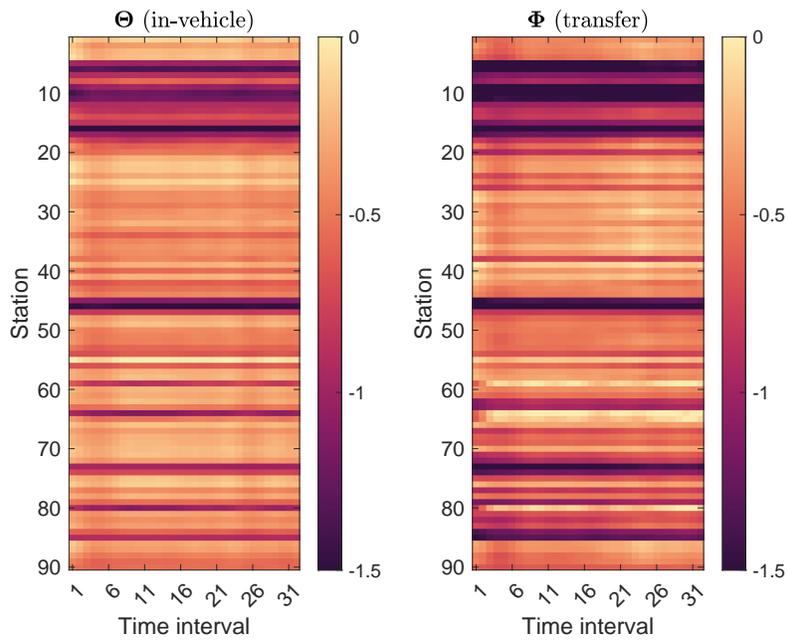

Fig. B.3: The upper bound of 95% credible interval of spatiotemporal choice parameters.

32

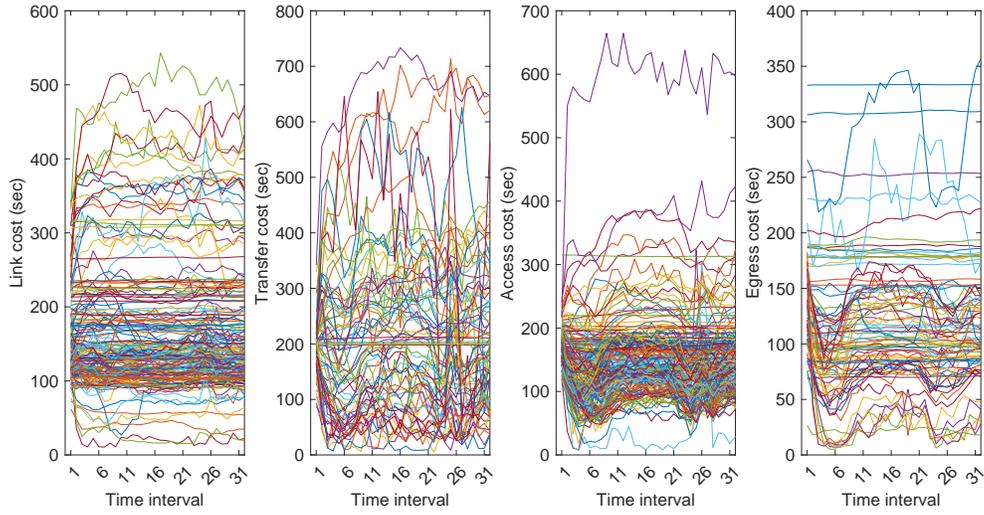

Fig. B.4: The estimated time-varying network cost parameters.

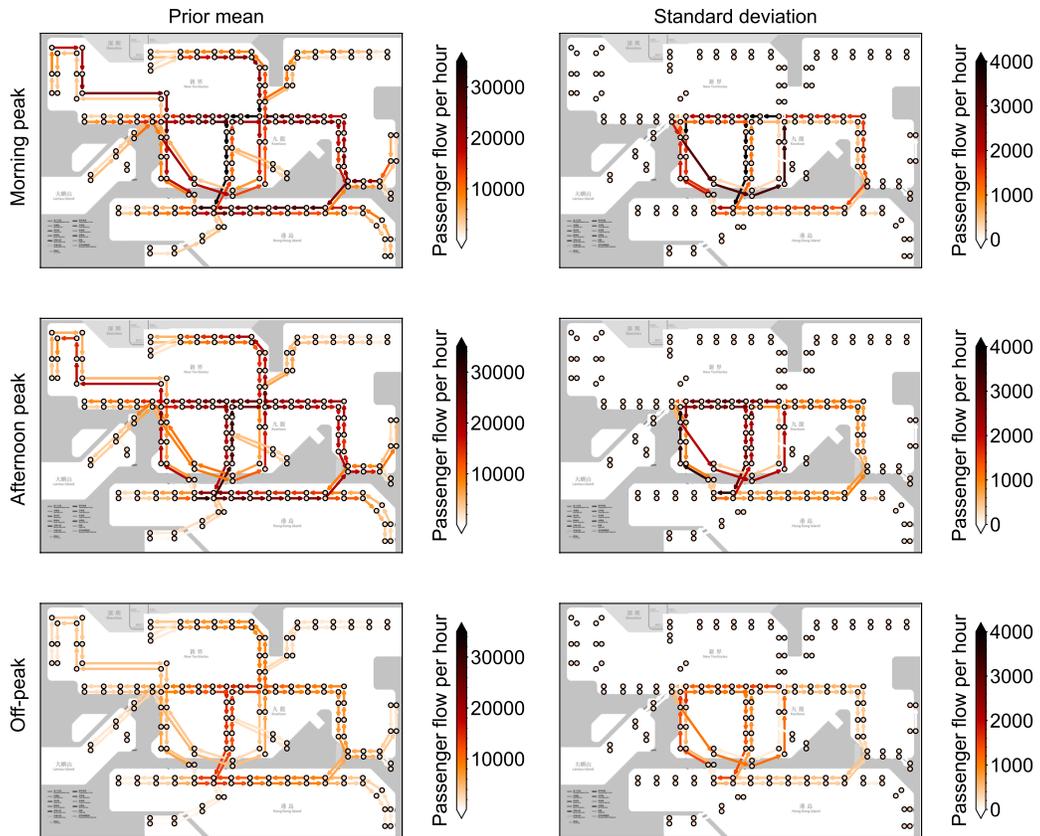

Fig. B.5: The prior means (first column) and pointwise standard deviations (second column) of passenger flow assignment under different times.

33



\end{document}